%% file: draft_qmm.tex
\documentclass[aps,prd,twocolumn,showpacs,superscriptaddress,superscriptaddress,nofootinbib]{revtex4-1}  

\usepackage{amsmath}
\usepackage{amsfonts}
\usepackage{amssymb}
\usepackage{braket}
\usepackage[separate-uncertainty=true]{siunitx}

\usepackage[colorlinks=true, urlcolor=blue, citecolor=blue, linkcolor=blue, filecolor=blue]{hyperref}

\usepackage{graphicx}     
\usepackage{multirow}
\usepackage{booktabs}

\newcommand*\diff{\mathop{}\!\mathrm{d}}
\DeclareMathOperator{\Tr}{Tr}
\DeclareMathOperator{\diag}{diag}
\DeclareMathOperator{\sgn}{sgn}
\usepackage{slashed}

\begin{document}
	
\title{Thermalization and dynamical spectral properties in the quark-meson model }
	
\author{Linda Shen}
\affiliation{Heidelberg University, Institute for Theoretical Physics, Philosophenweg 16, 69120 Heidelberg, Germany}
\author{J\"urgen Berges}
\affiliation{Heidelberg University, Institute for Theoretical Physics, Philosophenweg 16, 69120 Heidelberg, Germany}
\author{Jan M. Pawlowski}
\affiliation{Heidelberg University, Institute for Theoretical Physics, Philosophenweg 16, 69120 Heidelberg, Germany}
\author{Alexander Rothkopf}
\affiliation{University of Stavanger, Faculty of Science and Technology, Kristine Bonnevies vei 22, 4036 Stavanger, Norway}

\date{\today}
	
\begin{abstract}
	We investigate the nonequilibrium evolution of the quark-meson model using two-particle irreducible effective action techniques. Our numerical simulations, which include the full dynamics of the order parameter of chiral symmetry, show how the model thermalizes into different regions of its phase diagram. In particular, by studying quark and meson spectral functions, we shed light on the real-time dynamics approaching the crossover transition, revealing e.g.\ the emergence of light effective fermionic degrees of freedom in the infrared. At late times in the evolution, the fluctuation-dissipation relation emerges naturally among both meson and quark degrees of freedom, confirming that the simulation successfully approaches thermal equilibrium. 
	\end{abstract}
	
\maketitle
	
\section{Introduction}

The quest to discover the conjectured critical point of the QCD phase diagram is a central motivation of modern heavy-ion collision experiments at collider facilities, such as the Large Hadron Collider at CERN and the Relativistic Heavy-Ion Collider (RHIC) at Brookhaven National Laboratory. In the beam energy scan currently executed at RHIC, the phase diagram of QCD is explored over a wide range of temperatures and baryon densities by depositing different amounts of energy in the initial collision volume. As the fireball expands and cools, the efficient exchange of energy and momentum among quarks and gluons leads to local thermalization over time. The question to answer is: if a critical point exists and some of the volume of the fireball evolves close to it, does the dynamical buildup of long range fluctuations leave any discernible mark on the yields of measurable particles?

Understanding the out-of-equilibrium dynamics of heavy-ion collisions thus remains one of the most pressing theory challenges in heavy-ion physics. So far, genuinely nonperturbative \textit{ab-initio} calculations of the equilibration process of the quark-gluon plasma and the dynamics close to the phase transition remain out of reach. 
In order to make progress, we therefore set out to shed light onto pertinent aspects of the physics of dynamical thermalization in heavy-ion collisions by deploying a low-energy effective theory of QCD, the two-flavor quark-meson model. This model incorporates the off-shell dynamics of the lowest mass states in QCD, the pseudoscalar pions, the scalar sigma-mode, and the 
light up and down quarks. Further degrees of freedom, in particular the gluons, heavier quark flavors as well as higher mass hadronic resonances carry masses $\gtrsim 500$\,MeV and are neglected here. This low-energy effective theory reflects the central and physically relevant feature of low-energy QCD: chiral symmetry breaking in vacuum and its restoration at finite temperature and density. At its critical endpoint, the model is expected to lie in the same universality class as QCD and hence constitutes a viable low-energy effective theory to explore dynamical critical phenomena in QCD at finite temperature and density at scales $\lesssim 500$\,MeV. 

In the present work, we consider the real-time dynamics of the two-flavor quark-meson model with small current quark masses in a nonexpanding scenario; for progress on the out-of-equilibrium quark-meson model, see \cite{Berges:2002wr, Berges:2009bx, Berges:2010zv, Berges:2013oba}.  
In the presence of such an explicit chiral symmetry breaking, the equilibrium chiral transition at finite temperature is a crossover as confirmed for QCD at vanishing and small density; for recent results, see \cite{Bonati:2018nut, Borsanyi:2018grb, Ding:2019prx}. By the help of different initial conditions defined via the initial occupations of sigma and quark fields, we map out the thermalization dynamics for different regions of the phase diagram.  
This allows, for the first time, to fully study the thermalization dynamics including that of order parameters of chiral symmetry. An extension of the present study to the scenario of an expanding fireball should give access to the freeze-out physics of heavy-ion collisions. 

The evolution toward thermal equilibrium is viewed through the lens of the one- and two-point functions of the theory, which are computed with the two-particle irreducible (2PI) approach by means of their quantum equations of motion. These correlation functions not only provide complementary order parameters for the study of chiral symmetry restoration but also give direct access to the spectral properties, including the quasiparticle content of the system. Being genuine nonequilibrium quantities they map out the whole time evolution of the system including the physics of the crossover transition in the late-time limit. 

This paper is organized as follows. In Section~\ref{sec:model} we briefly review the quark-meson model and give an overview over our nonequilibrium and nonperturbative treatment. The numerical setup for the time evolution starting from free-field initial conditions quenched to a highly nonequilibrium environment is described. In Section~\ref{sec:spectra} we discuss the spectral functions of the bosonic and fermionic degrees of freedom, which provide information about the masses as well as the lifetimes of the dynamical degrees of freedom. We investigate the late-time limit of our simulations, which reveals the dynamical emergence of the fluctuation-dissipation relation and hence allows us to define a thermalization temperature. Finally, Section~\ref{sec:field} covers the results for the sigma field describing the order parameter of the quark-meson model. We further discuss the behavior of different order parameters in equilibrium which lead to a consistent pseudocritical temperature. 
In Section~\ref{sec:conclusion} we conclude with a summary. Appendix~\ref{app:eom} provides details about the evolution equations of the model including the relevant expressions for the deployed approximation scheme. 

\section{The quark-meson model}
	\label{sec:model}
	
QCD evolves from a theory of dynamical quarks and gluons at large momentum scales, the fundamental degrees of freedom, to a theory of dynamical hadrons at low momentum scales. This transition of the dynamical degrees of freedom is related to the mass gaps of the respective fields. It is by now well understood that the gluon degrees of freedom start to decouple at about 1\,GeV, that is above the chiral symmetry breaking scale $k_\chi$ of about 400\,MeV. 
Most of the hadron resonances are too heavy for taking part in the off-shell dynamics and we are left with the up, down and to some extend the strange quarks, as well as the pions and the scalar sigma mode; for details, see \cite{Pawlowski:2014aha, Fu:2019hdw}. Indeed, low-energy effective theories emerge naturally at low momentum scales from first principle QCD, and their systematic embedding leads us to the quark-meson model and its Polyakov loop enhanced version as QCD-assisted low-energy effective theories. While its quantitative validity has been proven for momentum scales $k$ with $k\lesssim 300$\,MeV \cite{Alkofer:2018guy}, it reproduces qualitative QCD features up to $k\lesssim 700$\,MeV. It is this natural QCD embedding as well as its robust QCD-type chiral properties that has triggered a plethora of works with the quark-meson model on the QCD phase structure with functional methods, see e.g.~\cite{Berges:1998sd, Schaefer:2004en, Skokov:2010wb, Herbst:2010rf, Kamikado:2012cp, Fu:2016tey}. 
More recently also real-time correlation functions in equilibrium have been investigated in e.g.\ \cite{Floerchinger:2011sc, Tripolt:2013jra, Pawlowski:2015mia, Jung:2016yxl, Yokota:2016tip, Pawlowski:2017gxj, Yokota:2017uzu, Wang:2017vis, Tripolt:2018qvi, Wang:2018osm, Jung:2019nnr}. 
	
(Pre-)Thermalization has been studied in the $O(N=4)$ symmetric scalar model coupled to fermions using a two-loop expansion to next-to-leading order of the 2PI effective action in \cite{Berges:2002wr,Berges:2004ce}. The model was studied extensively in Refs.~\cite{Berges:2009bx, Berges:2010zv} in the context of inflaton dynamics to describe nonequilibrium instabilities with fermion production from inflaton decay. In \cite{Berges:2013oba}, the model was investigated for highly occupied bosonic fields, where the predictions were shown to agree well with lattice simulation results in the classical-statistical regime. Further results 
for spectral functions in and out of equilibrium with 2PI effective action techniques can be found in \cite{Shen:2019jhl}, and with classical-statistical simulations in \cite{PineiroOrioli:2018hst,Boguslavski:2019ecc, Schlichting:2019tbr} for scalar theories, and in \cite{Boguslavski:2018beu} for Yang-Mills theory.
	
In this work, we build on these results and investigate the nonequilibrium evolution of the two-flavor quark-meson model: we consider two light quark flavors with isospin symmetry, up and down quarks with an identical current quark mass $m_{u/d}=m_\psi$, coupled to a scalar mesonic field $\sigma$ and a triplet of pseudoscalar “pions” $\pi^\alpha$ ($\alpha=1,2,3$) through a Yukawa coupling $g$. The classical action reads
\begin{widetext}
	\begin{align}
	S[ \bar{\psi}, \psi, \sigma, \pi]
	&= \int \diff ^4 x \, 
	\Big[
	\bar{\psi} \left(i \gamma^\mu \partial_\mu - m_\psi  \right) \psi 
	- \dfrac{g}{N_f} \bar{\psi} \left(\sigma + i \gamma_5 \tau^\alpha \pi^\alpha \right) \psi
	+ \frac{1}{2} \left( \partial_\mu \sigma \partial^\mu \sigma + \partial_\mu \pi^\alpha \partial^\mu \pi ^\alpha \right)
	\nonumber\\[1em]&\hspace{6cm}
	- \dfrac{1}{2} m^2 \left(\sigma^2 + \pi^\alpha \pi^\alpha\right)
	- \dfrac{\lambda}{4!N} \left(\sigma^2 + \pi^\alpha \pi^\alpha\right)^2
	\Big]\,,
	\label{eq:action}
	\end{align}
\end{widetext}
with $ \tau^\alpha $ ($ \alpha=1,2,3 $) denoting the Pauli and $ \gamma^\mu $ ($ \mu=0,1,2,3 $) the Dirac matrices, while spinor and flavor indices are suppressed. In \eqref{eq:action}, $m_\psi$ is the current quark mass and $ m^2 $ the mesonic mass parameter. The lowest mass states of the mesonic scalar-pseudoscalar multiplet, $\sigma$ and $\vec\pi$, are given by the $ N=4 $ scalar components of the bosonic field $ \varphi_a (x) = \{ \sigma(x), \pi^1(x), \pi^2(x), \pi^3(x)\} $ interacting via a quartic self-coupling $ \lambda $. 
The boson fields $ \varphi_a $ are coupled to the fermion fields $ \psi $ and $ \bar{\psi} = \psi ^ \dagger \gamma^0 $ via the Yukawa interaction $ g $, which we also express in terms of $ h = g/N_f $. 

The $ \pi $ mesons play the role of the light Goldstone bosons in the chirally broken phase whereas the $ \sigma $ meson represents the heavy mode. Assigning these roles to the components of the scalar field is achieved by choosing a coordinate system in field space where the field expectation value has a single component which defines the $ \sigma $ direction,  i.e. $ \phi_a(x) = \braket{\varphi(x)}=\{ \braket{\sigma(x)}, 0, 0, 0\} $. 

The quasiparticle excitation spectrum of the quark-meson model is encoded in the spectral functions of the respective fields. For the bosonic and fermionic fields, the spectral function is defined as the expectation value of the commutator and anticommutator, respectively, 
\begin{align}\nonumber 
	\rho^\phi_{ab}(x,y) 
	&=
	\  i \ \braket{[{\varphi}_a(x), {\varphi}_b(y)]}\,,\\[1ex]
	\rho^\psi_{AB}(x,y) 
	&= 
	\  i \ \braket{\{{\psi}_A(x), {\bar{\psi}}_B(y)\}}\,,
\label{eq:spectral_functions}	\end{align}
where $ a, b = 1, \dots, N$ denote field space  and $ A,B = 1, \dots, 4 $ correspond to Dirac spinor indices. Fermion flavor indices are omitted and the operator nature of the quantum fields is implied. 
We consider systems with spatial isotropy and homogeneity such that the spectral functions depend on times and relative spatial coordinates, i.e. $ \rho(t, t', |\mathbf{x} - \mathbf{y}|) $ or in momentum space $ \rho(t, t', |\mathbf{p}|) $, while the field expectation value only depends on time, i.e. $ \braket{\sigma(t)} $. 
Due to the remaining $ O(N-1) $ symmetry of the chirally broken model, the bosonic spectral function can be written as $ \rho^\phi_{ab} = \diag (\rho_\sigma, \rho_\pi, \rho_\pi, \rho_\pi)$ where the components $ \rho_i $ with $ i = \sigma, \pi $ describe the respective mesons. 
The fermionic spectral function can be decomposed into Lorentz components according to 
\begin{align}
\rho^\psi &=
\rho_S 	
+ i \gamma_5 \rho_P
+ \gamma_\mu \rho_V^\mu
+ \gamma_\mu \gamma_5 \rho_A^\mu
+ \frac{1}{2} \sigma_{\mu\nu} \rho_T^{\mu \nu}
\end{align}
with $ \sigma_{\mu\nu} = \frac{i}{2} \left[\gamma_\mu, \gamma_\nu \right] $ and $ \gamma_5 = i \gamma^0 \gamma^1\gamma^2\gamma^3 $. 
The corresponding \textit{Lorentz components} are given by 
\begin{align}\nonumber 
	\rho_S
	&= \frac{1}{4}\Tr \left[ \rho^\psi\right] \,, & 
	\rho_P
	&= \frac{1}{4}\Tr \left[ -i \gamma_5 \rho^\psi\right]
	\,, \\[1ex]\nonumber 
		\rho_V^\mu
	&= \frac{1}{4}\Tr \left[ \gamma^\mu \rho^\psi\right] 
		\,, & 
		\rho_A^\mu
	&= \frac{1}{4}\Tr \left[ \gamma_5 \gamma^\mu \rho^{\psi}\right]
	\,, \\[1ex]
	 \rho_T ^{\mu \nu}
	&= \frac{1}{4}\Tr \left[ \sigma^{\mu \nu}\rho^{\psi}\right]\,, 
\label{eq:Lorentz_components}\end{align}
where the trace acts in Dirac space. 
In spatially homogeneous and isotropic systems with parity and CP invariance, the only nonvanishing components are the scalar, vector and $ 0i $-tensor components. Rotational invariance allows us to write
	\begin{align}\nonumber 
	\rho_S(x^0, y^0, \mathbf{p}) 
	&= \rho_S(x^0, y^0, |\mathbf{p}|)\,, \\[1ex]\nonumber
	\rho_V^0 (x^0, y^0, \mathbf{p}) 
	&= \rho_0(x^0, y^0, |\mathbf{p}|)\,, \\[1ex]\nonumber
	\rho_V^i (x^0, y^0, \mathbf{p}) 
	&= \dfrac{p^i}{|\mathbf{p}|}\, \rho_V(x^0, y^0, |\mathbf{p}|)\,, \\[1ex]
	\rho_T^{0i} (x^0, y^0, \mathbf{p}) 
	&= \dfrac{p^i}{|\mathbf{p}|} \,\rho_T(x^0, y^0, |\mathbf{p}|)\,,
\label{eq:fermion_components}%
	\end{align}
where we refer to the two-point functions
$ \rho_S$, $\rho_0$, $ \rho_V$ and  $\rho_T $ on the right-hand sides as the scalar, vector, vector-zero and tensor components. The relevant contributions to the quark spectral function are the scalar, vector-zero and vector components, where the vector-zero component represents the quark excitations of the system \cite{Tripolt:2018qvi,Kitazawa:2006zi}. For chiral symmetric theories with $ m_\psi = 0 $ the scalar and tensor components vanish. 
The spectral functions also encode the equal-time commutation and anticommutation relations of the quantum theory, implying that 
\begin{align}
i \partial_t \rho^\phi(t, t', |\mathbf{p}|)\Big|_{t=t'} = 1\,, && 
\rho_0 (t,t, |\mathbf{p}| ) = i\,,
\label{eq:commutation_relations}
\end{align}
while all other fermion components vanish at equal time. 

In addition to the spectral functions, we may consider the so-called \textit{statistical functions}. These are the anticommutator and commutator expectation values,
	\begin{align}\nonumber 
	F^\phi(x,y) 
	&= 
	\frac{1}{2} \braket{\{\varphi(x),\varphi(y)\}} - \phi(x) \phi(y)\,,
\\[1ex]
	F^\psi(x,y) &= 
\frac{1}{2} \braket{[\psi(x),\bar{\psi}(y)]}\,,
\label{eq:statistical_functions}
	\end{align}
where field space, Dirac, and flavor indices are suppressed. The statistical functions carry information about the particle density of the system, i.e., the occupation of the available modes in the system. 
Together, the spectral and statistical functions fully describe the time-ordered connected two-point correlation function, commonly denoted as $G(x,y)=\braket{T \varphi(x)\varphi(y)}
-\braket{ \varphi(x)}
\braket{ \varphi(y)}$ for the bosonic and $\Delta(x,y)=\braket{T\psi(x)\bar{\psi}(y)}$ for the fermionic sector. Note that in nonequilibrium settings, the time-ordering occurs along the closed time path also known as Schwinger-Keldysh contour. 

\subsection{2PI effective action real-time formalism at NLO}

\begin{figure*}[t]
	\centering
	\includegraphics[width=.8\textwidth]{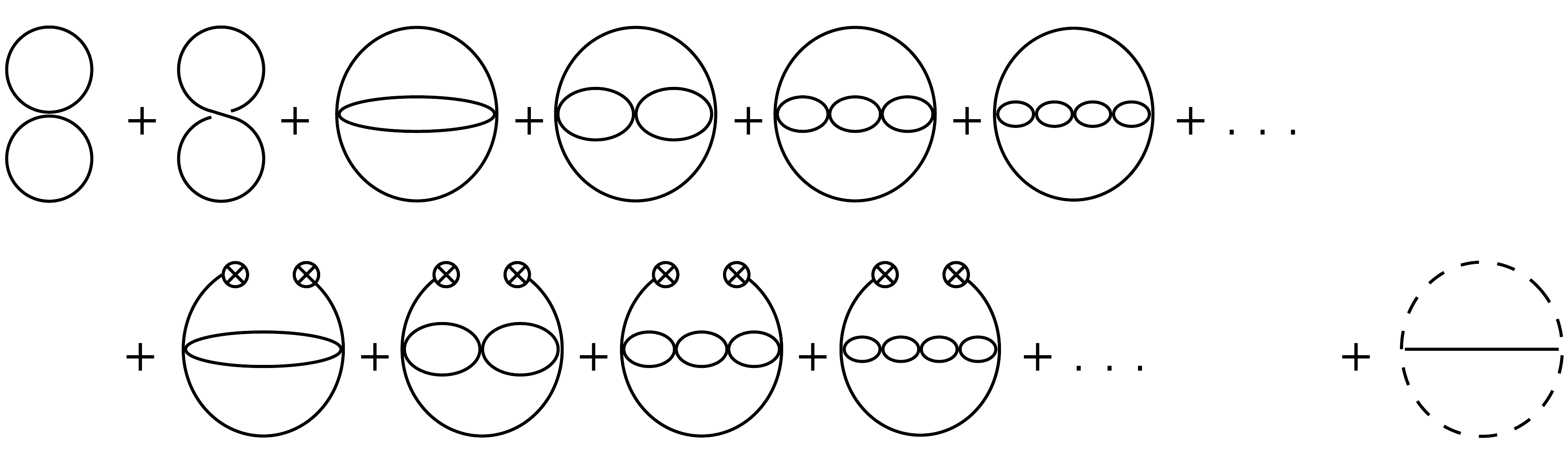}
	\caption{2PI diagrams at NLO in $ 1/N $ and $ g $. Full lines represent boson propagators, crossed circles macroscopic field insertions and dashed lines fermion propagators. 
		The first two-loop diagram in the first row corresponds to the leading-order contribution in $ 1/N $. The last diagram in the second row shows the fermion boson loop. The other diagrams in the first and second row depict the infinite series of NLO diagrams in $ 1/N $. 
	}
	\label{fig:Diagrams}
\end{figure*}

One can derive closed and nonsecular evolution equations for the one- and two-point functions of the quark-meson model out of equilibrium. These equations follow from the 2PI effective action $\Gamma[\phi,G,\Delta]$, the quantum counterpart of the classical action $S[ \bar{\psi}, \psi, \sigma, \pi]$, via a variational principle (see e.g. \cite{Berges2004}). 
The 2PI effective action of the quark-meson model can be written as 
\begin{align}
\Gamma[\phi, G, \Delta]
&= 
S[\phi]
+ \dfrac{i}{2} \Tr \ln  \left[G^{-1}\right ]
+ \dfrac{i}{2} 		\Tr \left[G^{-1}_{\text{cl}}(\phi) G\right ]
\nonumber\\[1em]& \quad
-i	\Tr \ln  \left[\Delta^{-1}\right ]
-i 			\Tr  \left[\Delta^{-1}_{\text{cl}}(\phi) \Delta\right ] 
\nonumber\\[1em]& \quad
+ \Gamma_2[\phi, G , \Delta] + \text{const}. \,, \label{eq:2PIeffectiveAction_QMM}
\end{align}
where $ S $ is the classical action given by \eqref{eq:action}, and $ G^{-1}_{\mathrm{cl}} $ and $ \Delta^{-1}_{\mathrm{cl}} $ are the classical meson and quark propagators derived from it. Traces, logarithms and products have to be evaluated in the functional sense. 
The term $ \Gamma_2[\phi, G, \Delta] $ contains two-loop and higher order quantum fluctuations that correspond to 2PI diagrams. 

The relevant evolution equations for the one- and two-point functions have the form (explicit expressions can be found in Appendix~\ref{app:eom}): 
\begin{align}\nonumber 
\left[
\square _x + M^2(x)
\right]
\phi(x) 
&= 
\int_{0}^{x^0}\!\!\!\! \diff z\,
\Sigma_\phi(x,z) \phi(z)
+ J_\phi(x)
\,,\\[1ex]\nonumber 
\left[
\square _x + M^2_\phi(x)
\right]
\rho^\phi(x,y) 
&= 
\int_{y^0}^{x^0} \diff z\,
\Sigma^\phi_\rho(x,z)
\rho^\phi(z,y)\,,\\[1ex]
\left[i \slashed{\partial}_x + M_\psi(x)
\right]
\rho^\psi(x,y) 
&= 
\int_{y^0}^{x^0} \diff z\,
\Sigma^\psi_\rho(x,z)
\rho^\psi(z,y)\,,
\label{eq:eom_sketch}%
\end{align}	
with shorthand notation $\int_{t_1}^{t_2}{\diff z} \equiv \int_{t_1}^{t_2}{\diff z^0 \int{\diff^3 z}}$ and the dependence of the self-energies $ \Sigma_i $ on $\phi,G, \Delta $ is implied. Similar expressions hold for the statistical functions. 
On the left-hand side the Klein-Gordon or Dirac operators act on the corresponding expectation value. Thereby, effective masses take into account local quantum corrections.
On the right, the effects of quantum fluctuations appear in so-called memory integrals that encode the generally non-Markovian effects of fluctuations in the past. The source term $ J _\phi$ in the field equation arises in the chirally broken case and describes the fermion backreaction on the field. It pushes the field to nonzero field expectation values even in the case where $ \phi(t) = \partial_t\phi(t)= 0 $ at initial time. 

In order to carry out explicit computations, the self-energies $\Sigma_i$ need to be approximated. Here we deploy an expansion to next-to-leading order (NLO) in $ 1/N $ for the bosons, where $ N $ is the number of scalar field components, and a NLO expansion in $ g $, the Yukawa coupling. The large $ N $ expansion provides a controlled nonperturbative approximation scheme, which at NLO includes scattering as well as off-shell and memory effects, capable of handling relatively large couplings \cite{Berges:2001fi}. 
The loop expansion in $ g $ to NLO contributes with a fermion-boson loop originally discussed in \cite{Berges:2002wr}. The 2PI diagrams contributing in this approximation are sketched in Figure~\ref{fig:Diagrams}. 

The explicit equations of motions are presented in Appendix~\ref{app:eom}, where also the self-energy expressions for the given approximation scheme are provided. To study the time evolution of the system, we iteratively solve the equations of motion without further approximations.

\subsection{Initial conditions}

The derivation of the nonequilibrium 2PI effective action and the equations of motion following from it rely on the assumption of a Gaussian initial state. This corresponds to a system initially exhibiting the characteristics of a noninteracting theory. However, higher order correlation functions build up during the subsequent time evolution. 
While this appears at first sight to correspond to a very limited choice of initial conditions, it still allows for a wide variety of different configurations through which we can determine for instance the energy density $\varepsilon_{\rm init}$ at the beginning of our computation. In particular, the Gaussian initial state represents a genuine nonequilibrium state in the fully interacting nonequilibrium system, in which the time evolution takes place. 

We allow for spontaneous symmetry breaking by using a negative mesonic bare mass squared $ m^2 <0$ in the classical potential of the system. Since the initial state is determined by a free theory with $ m^2 = m^2_\mathrm{init} > 0 $, the sign flip of $ m^2 $ leads to a quench of the classical potential from positive to negative curvature in the first time step. At initial time, the classical potential is minimal at vanishing field expectation value while the minimum at $ t>0 $ becomes nonzero by taking $ m^2 < 0 $.

A Gaussian initial state can be fully specified in terms of the one- and two-point functions. Since the field evolution equation involves second order time derivatives, one has to specify both the sigma field value and its initial time derivative. We choose the latter to vanish and refer to the initial field expectation value as $ \sigma_0 $,
\begin{align}
\sigma(t=0) = \sigma_0\,, && \partial_t \sigma(t) \Big|_{t=0} = 0\,,
\end{align}
where $ \sigma(t) $ now denotes the expectation value of the sigma field. 
As pointed out above, due to the presence of a finite bare quark mass $m_\psi$ the field can move away from $\sigma_0=0$ due to the backreaction with the fluctuations of the theory.

We specify the initial conditions for the two-point functions in terms of the spectral and statistical components. The initial conditions for the bosonic (fermionic) spectral functions are fully determined by the equal-time commutation and (anti)commutation relations \eqref{eq:commutation_relations}. 
For the remaining statistical functions we employ free-field expressions with a given initial particle number. 
The bosonic statistical function then reads
\begin{align}
F_i(t, t', |\mathbf{p}|  ) 
&= \dfrac{n_i(t, |\mathbf{p}|) + \frac{1}{2}}{\omega_i(t, |\mathbf{p}|) }
\cos\left[ \omega_i(t, |\mathbf{p}|)(t-t')\right]\,, 
\end{align}
with $  i = \sigma, \pi $ and where at initial time $ t = t' = 0 $ the dispersion is set to $  \omega_i(0, |\mathbf{p}|)  = \sqrt{|\mathbf{p}|^2 + m_\mathrm{init}^2}$ with initial mass squared $ m^2_\mathrm{init}>0 $ and the particle distribution given by $ n_i(0, |\mathbf{p}|) = 0 $. For the fermions the free statistical function can be written as
\begin{align}
F^\psi(t, t, |\mathbf{p}|) = \dfrac{- \gamma^i p_i + m_\psi}{\omega_\psi(t,|\mathbf{p}| )}
\left(
\frac{1}{2} - n_\psi(t, |\mathbf{p}|)
\right )\,, 
\end{align}
where we choose the initial dispersion to be $ \omega_\psi(0, |\mathbf{p}|) = \sqrt{|\mathbf{p}|^2 + m_\psi^2}$ and the initial particle distribution to be constant, i.e. $ n_\psi(0, |\mathbf{p}|) = n_0  $. 
 
 \begin{figure*}[t]
	\centering
	\includegraphics[width=0.6\textwidth]{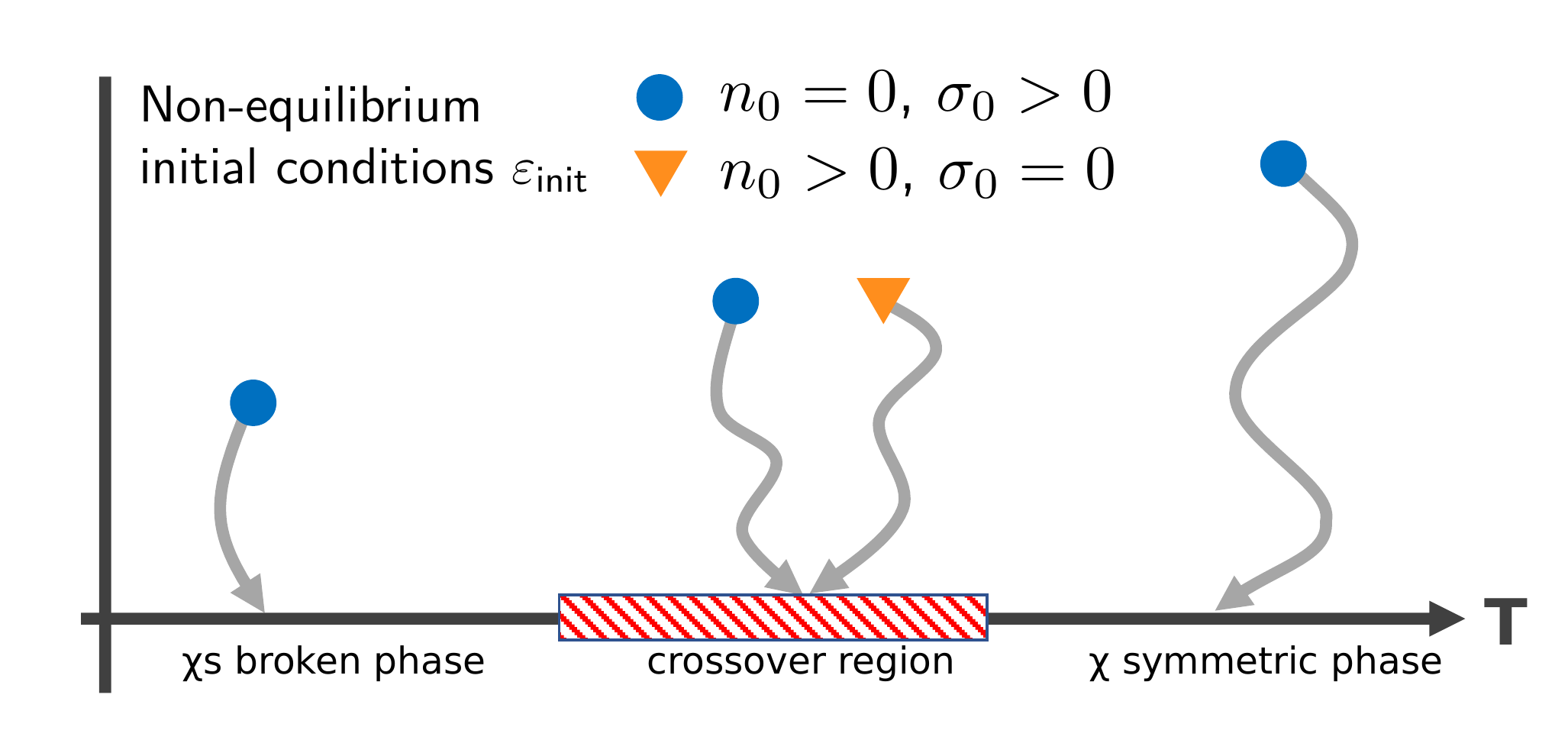}	
	\caption{Sketch of the setup deployed in this study. We consider the real-time evolution from nonequilibrium initial states characterized by an energy density sourced either through a finite $\sigma$ field expectation value (blue circle) or a nonzero occupancy of fermionic modes (orange triangle). Depending on the initial energy contained in the system, one of three discernible final states, the chiral broken phase, the crossover regime or the (almost) symmetric phase is approached.
	}
	\label{fig:Initial_setup}
\end{figure*}

The energy contained in the initial state via $\varepsilon_{\rm init}$ determines the temperature at which the system thermalizes. By preparing different initial conditions, we can study the thermalization process toward different temperatures and hence phases of the model as sketched in Figure~\ref{fig:Initial_setup}.

\subsection{Numerical implementation}
\label{sec:numerics_QMM}
As is customary in the context of the 2PI effective action, we discretize the system on the level of the equations of motion \eqref{eq:eom_sketch}. The explicit form of the equations allows us to deploy a leap-frog scheme, where in particular the fermionic two-point functions are discretized in a temporally staggered fashion. The two-point functions, as the name suggests, carry an explicit dependence on two temporal coordinates. Since the memory integrals contain the full time history, the required memory grows quadratically with the number of time steps.  In order to keep the computation manageable we reduce the memory burden by exploiting isotropy and homogeneity, which reduces the effective spatial dimensions to one. A modified Fourier transform based on Hankel functions allows us to evaluate the self-energy contributions in coordinate space and to simplify the convolutions in the memory integrals in momentum space. For this project we extended the code used in Ref.~\cite{Berges:2009bx} to include the additional nonvanishing fermionic two-point functions present in our setup; the source code for this project is available via the Zenodo repository under \cite{shen_linda_2020_3698136}.

In the spirit of effective field theories, we choose a UV cutoff at a high enough momentum scale. Below this scale, we consider quantum and statistical fluctuation within the 2PI framework. The ultraviolet parameters of our effective field theory are cutoff dependent and chosen such that physical observables, i.e., mass ratios and the pion decay constant, are reproduced.

The numerical time evolution is computed using a spatial grid with $ N_x = 200$ lattice points and a lattice spacing of $ a_x = 0.2 $. The time step size is chosen to be $ a_t = 0.05\, a_x$ guaranteeing energy conservation at the level of a few percent for the times analyzed. In the following all dimensionful quantities will be given in units of the pseudocritical temperature $ T_{pc} $, which has the value $ T_{pc} = 1.3\, a_x^{-1}$ determined according to the procedure described in Section~\ref{sec:equilibrium}; see Figure~\ref{fig:order_parameters}. Subsequently, we use $ m $ to denote the dimensionless ratio $ m / T_{pc} $ and likewise for all other dimensionful quantities. 

Interactions between the macroscopic field, the bosonic and the fermionic propagators lead to an exchange of energy between the different sectors. To observe an efficient energy exchange and equilibration process at computationally accessible times, it is necessary to study large couplings. We choose the quartic self-coupling $ \lambda = 90.0 $, the Yukawa coupling $ g=5.0 $, the bare mass squared $ m^2 = -0.0047 $ and the bare fermion mass $ m_\psi = 0.15 $. 
These parameters not only allow us to observe the equilibration of the system on time scales accessible computationally but also lead to reasonable values for the observables when compared to the phenomenological values known at $ T=0 $, where the pion decay constant is $ f_\pi \simeq \SI{93.5}{MeV} $, the meson masses are $ m_\sigma \simeq \SI{400}{MeV} $ and $ m_\pi \simeq \SI{135}{MeV} $, and the constituent quark mass is $ m_q = \SI{350}{MeV} $ \cite{Tanabashi:2018oca}. 

The above choices are close to that used in equilibrium computations of the quark-meson model with functional methods and a physical ultraviolet cutoff $\Lambda_{\textrm{UV}}\approx 1$\,GeV, (see e.g. \cite{Cyrol:2017ewj, Fu:2019hdw}). In these computations it can be shown that the self-interaction is of subleading relevance for the fluctuation dynamics, despite the large size of the classical coupling $\lambda$. 
In the present 2PI framework, the quantum interactions are obtained through an NLO resummation and for large occupancies or large classical coupling they can be shown to be small. 

The functional equilibrium studies \cite{Cyrol:2017ewj, Fu:2019hdw}, as well as a comparison of the quark-meson model to QCD, (see e.g. \cite{Alkofer:2018guy}) reveal that a one-to-one correspondence of the low-energy limits of both theories in quantitative approximations to the full dynamics in the quark-meson model either requires a far smaller UV cutoff for the latter or a systematic improvement of the model towards QCD-assisted low-energy effective theories \cite{Pawlowski:2014aha, Fu:2019hdw}. In the present work, we restrict ourselves to studying the qualitative properties of the nonequilibrium dynamics as a first step.

 When identifying the sigma field expectation as pion decay constant, we can reproduce $ f_\pi < m_\pi < m_q < m_\sigma $ at low temperatures. At the lowest temperatures considered in this work, we find $ f_\pi / m_\pi \simeq  0.65$, which is very close to the phenomenologically known value of approximately $ 0.69 $, the meson mass ratio $ m_\sigma / m_\pi \simeq  1.75$, smaller than the vacuum value of around $  2.9$ but expected to increase when going to lower temperatures, and $ m_q / m_\pi =  1.45$, being on the order of magnitude with zero-temperature value of $ 2.6 $. Hence we expect our findings to qualitatively reproduce the QCD dynamics. Note however, that the meson mass ratio $ m_\sigma / m_\pi < 2$ leads to another order of the thresholds for scattering processes, and hence respective difference in the spectral functions.   

For the bosonic sector, we use vacuum initial conditions, i.e. $ n_\phi(t=0, |\mathbf{p}|) = 0 $. The initial mass is fixed at $ m^2_\mathrm{init} = 0.0047 $. The fermion initial distribution is chosen to be constant $ n_\psi(t=0, |\mathbf{p}|) = n_0 $. We study simulations with fluctuation dominated initial conditions where the fermion number $ n_0 $ is varied between $ 0 $ and $ 1$ while the initial field value is $ \sigma_0 = 0 $. 
Furthermore, the field dominated initial conditions with a nonvanishing field value of $ \sigma(t=0) = \sigma_0 $ between 0 and $ 2.0 $ with vanishing fermion number $ n_0 = 0 $ are investigated. Unless otherwise specified, plots are shown for the case $  n_0 = 0$ and $ \sigma_0 = 0 $. For plots showing spectral and statistical functions in frequency space a cubic spline interpolation of the data points is employed.

\section{Spectral functions}
\label{sec:spectra}

In this section, we explore the nonequilibrium evolution of the quark-meson model from the point of view of its quark and meson spectral functions. As these quantities are derived from the two-point correlation functions, they provide insight on the (quasi)particle content of the theory, the dispersion relation of propagating modes and their decay widths, providing insight into the modification of the system due to the presence of a (non)equilibrium medium. Our numerical simulations find clear indications for quasiparticles in both the IR and UV, revealing the presence of additional light propagating fermion modes for temperatures above the pseudocritical temperature. 

It is convenient to analyze the spectral functions in the \textit{Wigner representation} where the Fourier transformed spectral function can be interpreted as the density of states such that its structure provides information about the quasiparticle states of the system. 
Therefore, the temporal dependence of the unequal-time two-point correlation functions on the two times $ t $ and $ t' $ is rephrased in terms of Wigner coordinates: the central time $ \tau = (t + t')/2 $ and the relative time $ \Delta t = t-t' $. The dynamics in $ \Delta t $ describes microscopic properties of the system while the evolution in $ \tau $ describes macroscopic properties governed by nonequilibrium characteristics of the system. 
In order to study the frequency spectrum of the spectral functions, we then apply a \textit{Wigner transformation} to the propagators. 
This corresponds to a finite range Fourier transformation of the propagators with respect to the relative time $\Delta t $, which is constrained by $ \pm 2\tau$ in initial value problems where $ t, t' \geq 0 $. As a result, we obtain the frequency space spectral function
\begin{align}
\rho(\tau, \omega, |\mathbf{p}|) 
&= \int  _{-2\tau}^{2\tau}\mathrm{d} \Delta t \ 
e^{i\omega \Delta t} 
\rho \left( \tau, \Delta t, |\mathbf{p}|\right )\,,
\label{eq:Wigner_transformation}%
\end{align}
with analogous expressions for all statistical functions. 
For a real and antisymmetric spectral function (as in the bosonic case and for the fermionic scalar, vector and tensor components) as well as for an imaginary and symmetric spectral function (as for the fermionic vector-zero component), the Wigner transform $ \rho(\tau, \omega, |\mathbf{p}|) $ is imaginary. Due to symmetry, it is sufficient to present the Wigner transformed spectral functions for positive frequencies $ \omega $. Since the frequency space spectral functions are imaginary in our definition, we always plot $ -i \rho $ in the subsequent sections, thereby omitting the $ -i $ in the plot labels to ease notation.

The commutation and anticommutation relations \eqref{eq:commutation_relations} can be rephrased in frequency space, 
\begin{align}
\int \dfrac{\diff \omega}{2\pi}\  \omega\, \rho^\phi(\tau,\omega, |\mathbf{p}|) =i\,, &&\int \dfrac{\diff \omega}{2\pi}\ \rho_0(\tau, \omega, |\mathbf{p}|) = i\,,\label{eq:sum_rules}
\end{align}
where they are referred to as \textit{sum rules}. 
In our numerical computations, the bosonic and fermionic sum rules are satisfied at the level of 
$ \mathcal{O}(10^{-2}) $ and $ \mathcal{O}(10^{-6}) $, respectively. 

\subsection{Establishing thermal equilibrium at late times}
\label{sec:thermal_equilibrium}

Before embarking on a detailed study of the dynamical approach to thermal equilibrium, we first ascertain that our simulations of the quark-meson model exhibit thermalization at late times. We do so by observing the dynamic emergence of the fluctuation-dissipation theorem. One needs to keep in mind that as discussed in \cite{Berges:2002wr}, the idealized thermal equilibrium state cannot be reached in principle due to the time reversibility of the evolution equations. The simulation approaches the state more and more closely over time and at some point becomes indistinguishable from it for a given resolution. Hence, we expect the computation to approach a steady state. 

The fluctuation-dissipation theorem is reflected in a particular property of the spectral and statistical functions in thermal equilibrium: they are not independent of each other. In four-dimensional Fourier space, it reads
	\begin{align}\nonumber 
	F^\phi_{\text{eq}}(\omega, \mathbf{p}) 
	&=-i \left(
	\frac{1}{2} + n_{\text{BE}} (\omega) 
	\right )\rho^\phi_{\text{eq}}(\omega,\mathbf{p})\,,\\[1ex]
	F^\psi_{\text{eq}}(\omega, \mathbf{p}) 
	&=-i \left(
	\frac{1}{2} - n_{\text{FD}} (\omega) 
	\right )\rho^\psi_{\text{eq}}(\omega,\mathbf{p})\,,
\label{eq:FDT}%
	\end{align}
with $ n_{\text{BE}} (\omega)  = (e^{\beta\omega} - 1)^{-1}$ being the Bose-Einstein and $ n_{\text{FD}} (\omega)  = (e^{\beta\omega} + 1)^{-1}$ the Fermi-Dirac distribution. In \eqref{eq:FDT}, the frequency $ \omega $ is the Fourier conjugate to the relative time $ \Delta t = t-t' $ as the time dependence of $ F_\mathrm{eq} $ and $ \rho _\mathrm{eq}$ can be fully described in terms of $ \Delta t$ due to the time-translation invariance of thermal equilibrium. 

\begin{figure*}
	\centering
	\includegraphics[width=1.\textwidth]{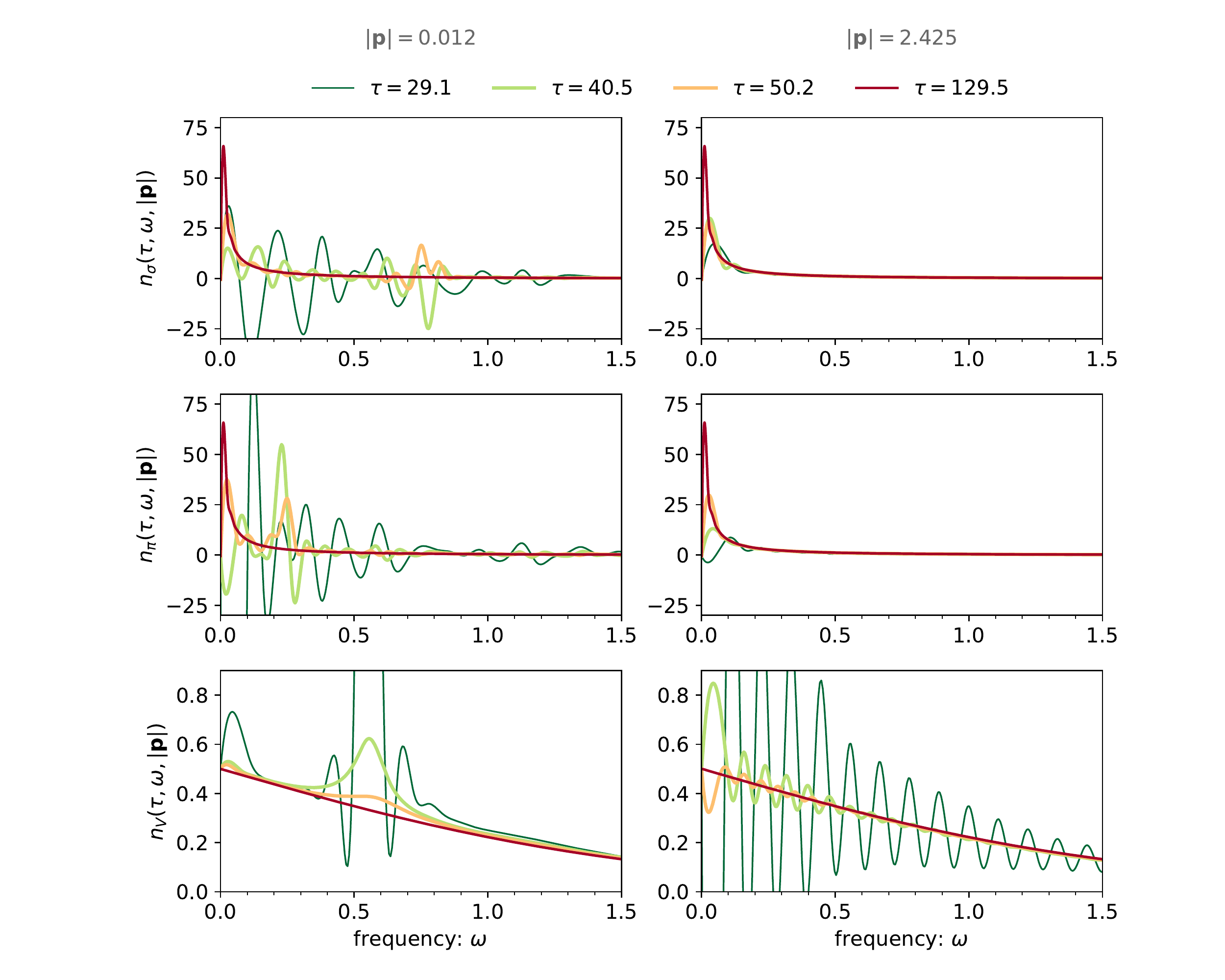}
	
	\caption{
		We show the time evolution of the effective particle number defined in \eqref{eq:n_eff} for bosonic and fermionic components (rows) and two different momenta (left and right columns). 
		At late times (red curve), the effective partcle number becomes time and momentum independent and approaches the shape of a Bose-Einstein and Fermi-Dirac distribution, respectively. The shown data are interpolated using a cubic spline. 
		Dimensionful quantities are given in units of the pseudocritical temperature $ T_{pc} $ (cf.  Figure~\ref{fig:order_parameters}).
		}
	\label{fig:Evolution_FDT}
\end{figure*}

Out of equilibrium, the independence of $ F $ and $ \rho $ manifests itself in the fact that the ratio $ F/\rho $ in general carries a momentum dependence. The equilibrium relation \eqref{eq:FDT} on the other hand allows us to define the generalized particle distribution function \cite{Berges2004}
\begin{align}
n_i (\tau, \omega, |\mathbf{p}|)&=\pm\left(
 i\  \dfrac{F_i(\tau, \omega, |\mathbf{p}| )}{\rho_i(\tau, \omega, |\mathbf{p}|)} - \dfrac{1}{2}
\right)
\,, 
\label{eq:n_eff}
\end{align}
with a positive (negative) sign for bosonic (fermionic) components and $ i = \sigma, \pi, V $. This kind of distribution function has been studied in the context of nonthermal fixed points in relativistic as well as nonrelativistic scalar field theories \cite{Orioli:2015dxa}.  
Considering \eqref{eq:n_eff} the approach of thermal equilibrium in a general nonequilibrium time evolution setup is characterized by $ n_i (\tau, \omega, |\mathbf{p}|)\rightarrow n_\mathrm{BE/FD}(\omega) $. 

In Figure~\ref{fig:Evolution_FDT} we show the time evolution of the particle distribution defined in \eqref{eq:n_eff} for low and high momenta (left and right columns). One can see that at late times (red curves) the same shape is approached for small and large momenta, whereas at early times the distribution functions differ from each other. This loss of momentum dependence is required for the thermalization process and reflects the emergence of the fluctuation-dissipation relation in the equilibrium state. From the late-time distributions shown in Figure~\ref{fig:Evolution_FDT} one can already guess that thermal distribution functions are reached.   

We also observe that the evolution of the effective particle number is different for fermions and bosons. The bosonic distribution functions $ n_\sigma $ and $ n_\pi $ show strong oscillations along frequencies at low momenta whereas oscillations at high momenta are weak. Since the particle distributions are computed by taking the ratio of the statistical and spectral functions, $ n_i $ plotted against $ \omega $ essentially describes how similar the peaks shapes of $ F $ and $ \rho $ are. In the high-momentum range we find that the quasiparticle peaks of the bosonic statistical and spectral functions resemble one another from early times on, while in the low-momentum range more time is required for the peak shapes to become aligned. In contrast, the quarks show an opposite behavior. Their distributions have much stronger frequency oscillations for large momenta than for small momenta, i.e. it takes longer for the high-momentum modes   to approach a thermal distribution. 

Putting the pieces together, we can see that a redistribution of the occupancies in fermionic and bosonic degrees of freedom occurs during the nonequilibrium time evolution. While the time scales to converge to thermal distribution functions depend on the particle species and the momentum modes we find that the distribution functions all become stationary for times $ \tau \gtrsim 100$, reflecting the time-translation invariant property of thermal equilibrium.

 \begin{figure*}[t]
 	\centering
 	\includegraphics[width=1.\textwidth]{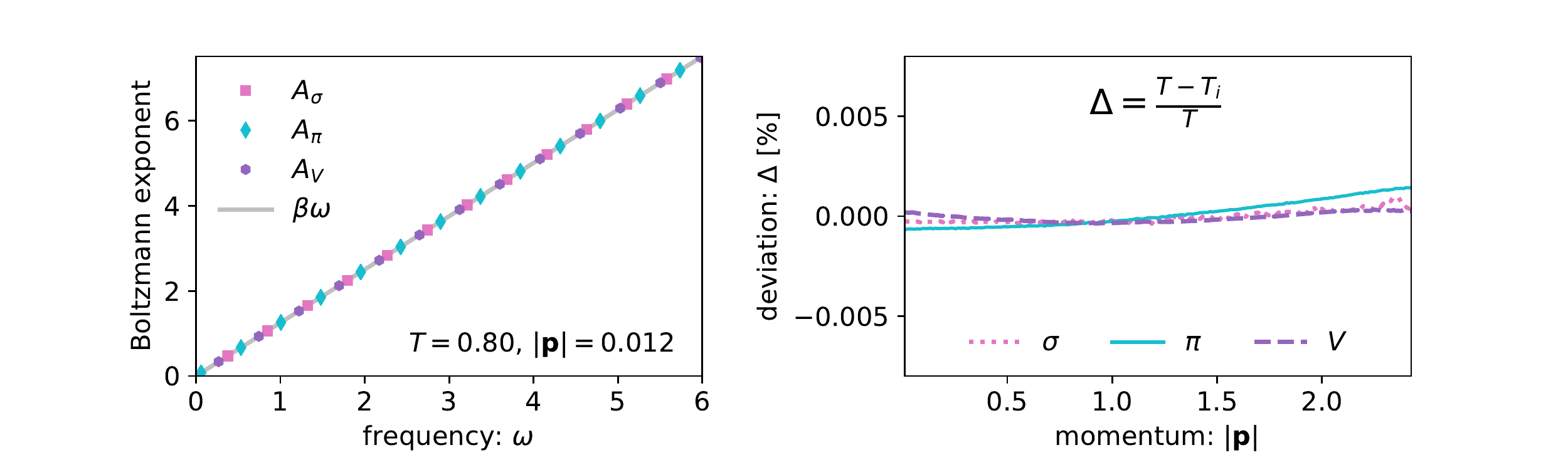}	
 	\caption{
 		Left: The generalized Boltzmann exponents defined in \eqref{eq:thermalization_temperature} shown as a function of frequency $ \omega $ at a given momentum $ |\mathbf{p}| $ for bosonic and fermionic components. For better visibility, only every 39th data point is shown. Using a linear fit one can determine the slope $ \beta $ and hence the temperature $ T $ for each component. The temperature $ T $ indicated in the plot is averaged over all momenta and the three components.\\
 		Right: The relative deviation from the thermalization temperature $  \Delta = (T_i - T) /  T $ shown for all three components as a function of momentum. The results for the bosonic and fermionic sectors agree very well. \\
 		In both plots dimensionful quantities are given in units of the pseudocritical temperature $ T_{pc} $ (cf.  Figure~\ref{fig:order_parameters}).
 	}
 	\label{fig:Thermalization}
 \end{figure*}
 
Although Figure~\ref{fig:Evolution_FDT} already indicates the approach of thermal distribution functions, we still need to prove whether our final state actually fulfils the fluctuation-dissipation theorem. 
For a quantitative analysis, we compute the \textit{generalized Boltzmann exponents}
\begin{align}
A_i(\tau, \omega, |\mathbf{p}|) = \ln \big[
	n_i^{-1}(\tau, \omega, |\mathbf{p}|) \pm 1
\big]\,, 
\label{eq:thermalization_temperature}
\end{align}
with positive (negative) sign for bosonic (fermionic) components and $ i = \sigma, \pi, V $. In thermal equilibrium, the fluctuation-dissipation theorem \eqref{eq:FDT} requires these exponents to suffice $ A_i(\tau, \omega, |\mathbf{p}|) = \beta \omega $, implying in particular that they become independent of momentum $ |\mathbf{p}| $ and time $ \tau $, where the latter is fulfilled by our late-time states. 

A linear fit of our simulation data for the generalized Boltzmann exponents to $ \beta \omega $ yields the thermalization temperature $ T_\mathbf{p} = \beta^{-1}_\mathbf{p}$,  which can in general be $ \tau $ dependent. An example for such a fit is presented on the left side of Figure~\ref{fig:Thermalization}. The plot shows that the Boltzmann exponent of all three components $ i = \sigma, \pi, V $ nicely fits to the same line with slope $ \beta $. We compute the temperature averaged over all momenta to obtain $ T_i $ for each component. The system temperature denoted by $ T $ is taken to be the mean over all three components. 

For every simulation, we compute the temperatures at each momentum $ |\mathbf{p}| $ and study the momentum dependence of the obtained temperature $ T_\mathbf{p} $. 
As pointed out in \cite{Berges:2004ce}, thermodynamic relations can become valid before real thermal equilibrium is attained, a phenomenon known as \textit{prethermalization}. Thermal equilibrium is characterized by $ T_\mathbf{p} $ being equal to some equilibrium temperature for all modes $ |\mathbf{p} |$. 
On the right side of Figure~\ref{fig:Thermalization}, the deviations from the mean thermalization temperature $ T $ are plotted. As can be seen, the deviations are very small. Hence, the Boltzmann exponents at late times $ \tau $ become momentum independent and the late-time states are thermal in the sense that they fulfill the fluctuation-dissipation theorem. The thermalization temperatures for all simulations in this work have been determined at time $ \tau = 130$. For the example shown in Figure~\ref{fig:Thermalization} it was checked that the thermalization temperatures found in the time range between $ \tau = 100 $ and $ \tau = 160 $ are constant at the level of $\mathcal{O}(10^{-3}) $. We have checked for all simulations in this work that the temperature has reached a stationary value at time $ \tau = 130 $.

Having clarified the successful approach to quantum thermal equilibrium in our system, we are now able to study the differences during the out-of-equilibrium evolution leading to the thermal states in detail. 

\subsection{Nonequilibrium time evolution of the spectral and statistical functions}
\label{eq:noneq_evol_2PF}

In this section we study the dynamics of the thermalization process, starting from fluctuation or field dominated initial conditions. We investigate the time evolution of the spectral and statistical functions and consider derived quantities such as particle masses and widths. While the initial conditions strongly influence the nonequilibrium dynamics taking place, the final states are universal and characterized by the initial energy density ${\varepsilon_{\rm init}}$ that translates into a unique temperature. 

The time evolution leads to the emergence of quasiparticle peaks in the spectral functions of both quark and mesons. The value of the particle mass and its decay width are a consequence of the interactions taking place among the microscopic degrees of freedom. While the initial states correspond to free particles, which would have a spectrum given by a $ \delta $ distribution located at the mass parameters of the classical action, the scattering effects included in the nonequilibrium evolution lead to peaks with finite widths in the spectrum. 

\begin{figure*}[t]
	\centering
	\includegraphics[width=1.\textwidth]{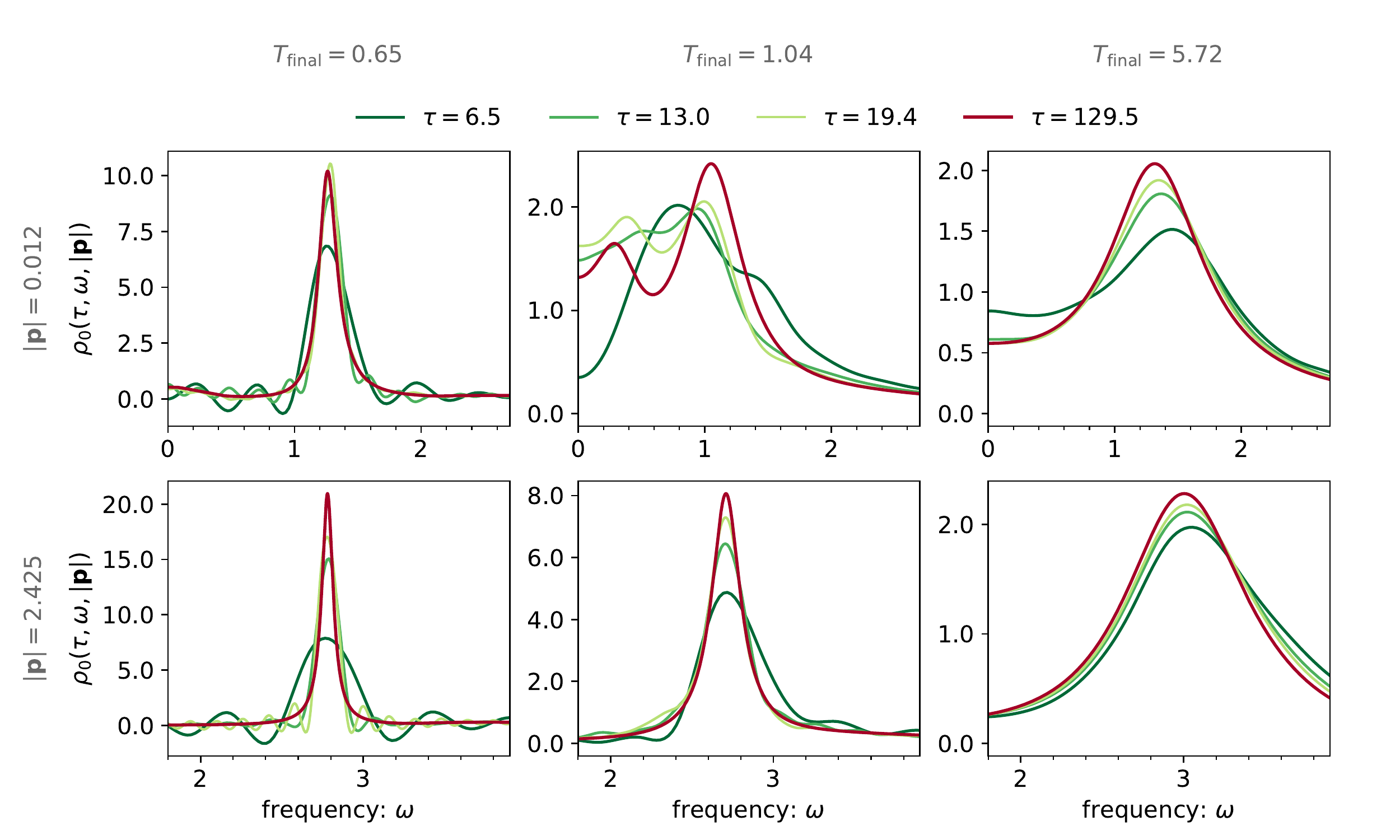}	
	\caption{A representative selection of spectral functions from the fermion vector-zero channel in the infrared (top row) and the UV (bottom row) in three different regimes labeled by the temperatures of their final state. Each panel contains four curves indicating different snapshots along the thermalization trajectory. All three simulations employ field dominated initial conditions, i.e. $ \sigma_0 >0 $ and $ n_0 = 0 $. Dimensionful quantities are given in units of the pseudocritical temperature $ T_{pc} $ (cf.  Figure~\ref{fig:order_parameters}).}
	\label{fig:Spectra_evolution_summary}
\end{figure*}

In Figure~\ref{fig:Spectra_evolution_summary} we present a representative set of fermionic spectral functions from the vector-zero channel, which describes the quark excitation spectrum \cite{tuprints2209, Kitazawa:2006zi}. The three columns correspond to three different field dominated initial conditions of increasing initial energy density, as sketched by the blue dots in Figure~\ref{fig:Initial_setup}. The top row shows the Wigner space spectral function at the lowest available momentum (IR), the bottom row at the highest momentum (UV). We can identify several characteristic properties of these spectral functions from a simple inspection by eye.

In the UV, a single quasiparticle structure is present at all times and at all energy densities. With increasing energy density in the initial state, corresponding to an increasing final temperature, the position of the peak and its width increase. This is consistent with the expectation that a fermion in an energetic medium will be imbued with an in-medium mass (to lowest order in perturbation theory it would be proportional to the temperature). Higher energy densities go hand in hand with an increased chance of scattering between the fermion and the other medium constituents, which also leads to a larger in-medium width. In the UV, no qualitative difference exists between the broken, crossover, or symmetric phase behavior. 

On the other hand, in the IR a clear distinction between the crossover region and all other energy density regimes is visible. While we also find a single quasiparticle structure at low and high initial energy densities, in the crossover region at early times no well-defined peaks are present at all. Instead, as times passes, two structures emerge. One dominant peak is located where one would expect the usual quasiparticle excitation to reside, another peak sits close to the frequency origin, denoting a significantly lighter additional propagating mode.

In general, we find that also for the other fermionic and bosonic spectral functions the approach of the equilibrium state depends on the initial conditions. In the presence of a nonzero initial field value $ \sigma_0 $, the spectral functions evolve differently than in the case where $ \sigma_0 = 0 $ but the fermion occupation is finite, i.e. when the initial state contains more energy in terms of fermion occupations. As pointed out in Figure~\ref{fig:Spectra_evolution_summary}, the most interesting dynamical features can be seen in the low-momentum area, which we therefore focus on during the following analysis. 

\subsubsection{High energy densities}
\label{sec:high_energy_densities}
Here, we study the quark-meson model at high enough initial energy densities such that the late-time evolution thermalizes in the high-temperature phase, where chiral symmetry is restored. 
For our analysis we compare two simulations starting from different initial conditions characterized by almost indistinguishable energy densities. One is dominated by the field $ \sigma_0 = 1.36 $ and $ n_0 = 0 $, while the other is dominated by fermion fluctuations $ \sigma_0 = 0 $ and $ n_0 = 0.8 $.
The final states feature similar thermalization temperatures of $ T = 3.15 $ and $ T =3.18$, respectively. However, since the initial states are very different from each other, the evolution toward thermal equilibrium takes significantly different paths. 
\begin{figure*}[t]
	\centering
	\includegraphics[width=1.\textwidth]{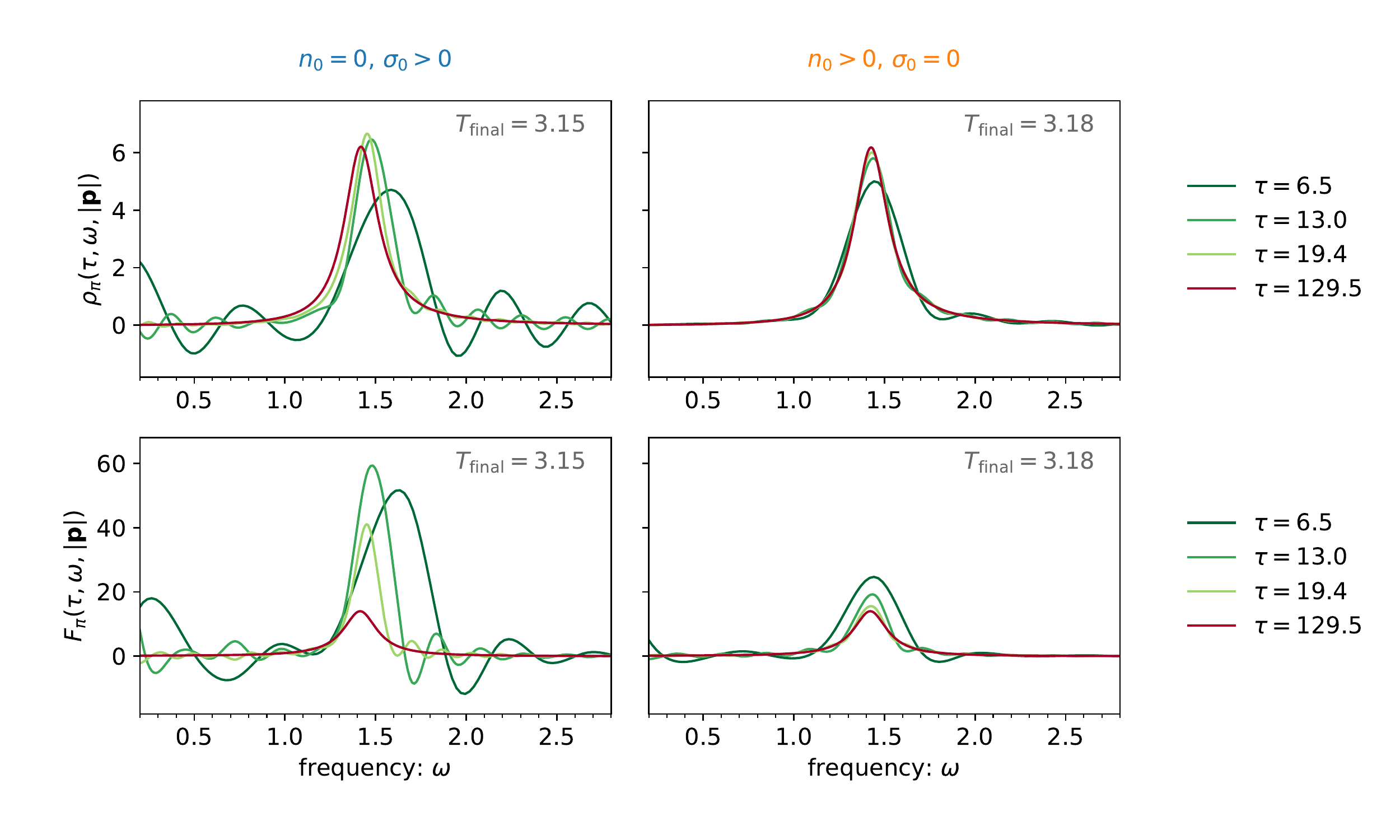}	
	\caption{
		Time evolution of the pion spectral and statistical functions shown for two different initial conditions at the smallest available momentum $ |\mathbf{
			p}| = 0.012 $. 
		The left column shows a simulation deploying field dominated initial conditions with $ \sigma_0 = 1.36$, 
		the right column fluctuation dominated initial conditions with $ n_0 = 0.8 $. 
		Both simulations lead to thermal states at temperatures where chiral symmetry is restored. 
		Dimensionful quantities are given in units of the pseudocritical temperature $ T_{pc} $ (cf.  Figure~\ref{fig:order_parameters}).
	}
	\label{fig:Spectra_evolution_highT_pion}
\end{figure*}

For such high initial energy densities, the differences in the time evolution are most apparent in the bosonic sector. This can be studied by looking at the bosonic spectral and statistical functions. Numerical results are shown in Figure~\ref{fig:Spectra_evolution_highT_pion}, where only the pion spectral and statistical functions are presented since the behavior of the sigma meson is analogous. 
The final states of both simulations (red curve) are characterized by the same peak shapes for both spectral and statistical functions. However, the functions at intermediate times exhibit a completely different behavior.

For field dominated initial conditions (left column in Figure~\ref{fig:Spectra_evolution_highT_pion}), the peak position of the spectral function moves toward smaller frequencies with time, which means that the mass of the quasiparticle state decreases during the time evolution. In addition, the nonzero initial field leads to large amplitudes in the pion statistical function at early times (lower left plot in Figure~\ref{fig:Spectra_evolution_highT_pion}) which corresponds to relatively high occupancies in the bosonic sector compared to the final thermal distribution. These occupancies have to redistribute to other bosonic momentum modes $ |\mathbf{p}| $ and the fermionic sector to let the system equilibrate. 

This behavior can be readily understood from the microscopic evolution equations of the system. The finite-valued initial field drives the fluctuations in the bosonic sector because it contributes to the bosonic self-energy at initial time $ t=0 $. Since the nonequilibrium time evolution takes into account the full time history since $ t=0 $, these initial fluctuations not only play a role at initial time but also at intermediate times. 
Only at late times, the system loses memory about the details of the initial state. Since the macroscopic field only couples to the bosons directly but not to the fermions, the energy provided by the initial field is first turned into bosonic fluctuations before being transferred to fermionic modes. As a consequence, the thermalization of an initial state with nonzero initial field value shows rich dynamics in the bosonic spectra. 

In contrast, for fluctuation dominated initial conditions (right column in Figure~\ref{fig:Spectra_evolution_highT_pion}), one observes a continuous increase of the amplitudes of both spectral and statistical functions until the maximum is reached in the thermal state. If the initial energy density is provided via fermionic fluctuations, the thermal final state is found to be realized already at intermediate times. 

\begin{figure*}[t]
	\centering
	\includegraphics[width=1.\textwidth]{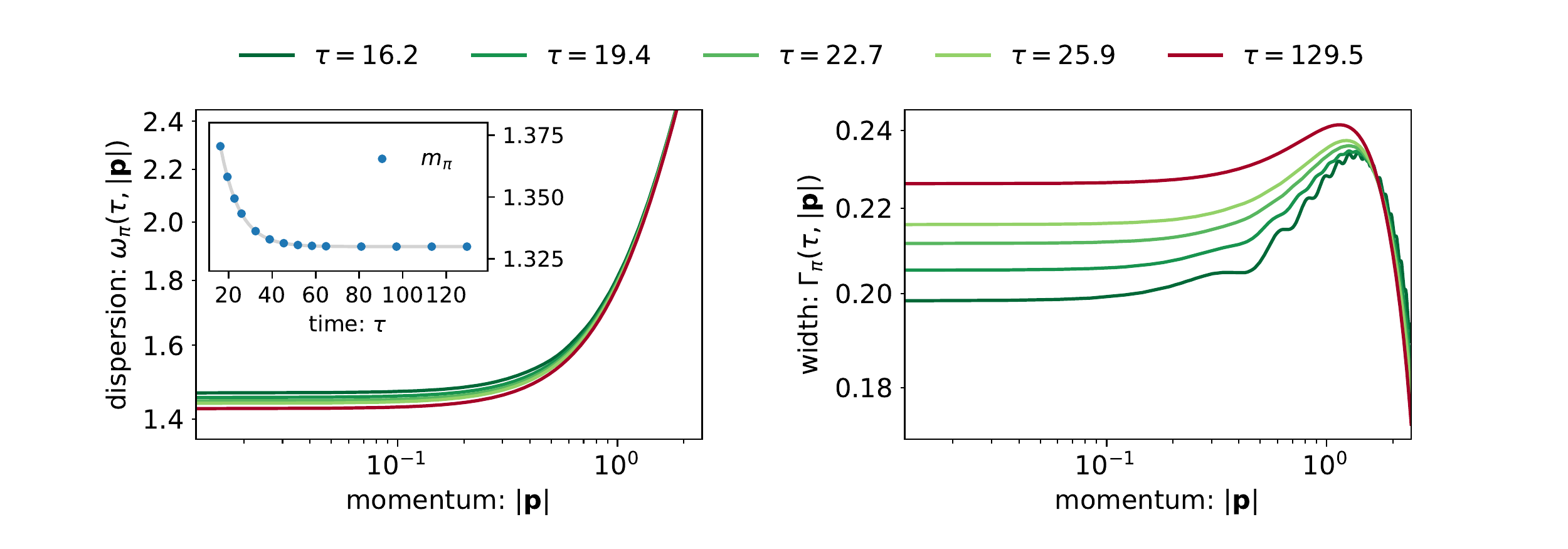}	
	\caption{
		Time evolution of the dispersion relation and the momentum dependent width of the pion. The inset shows the time evolution of pion mass obtained from fits of the dispersion relation to $ Z \sqrt{|\mathbf{p}|^2 + m_\pi^2} $ at various times $ \tau $, where $ Z = 1.07 $ is obtained for all times analyzed. The data are shown for field dominated initial conditions with $ \sigma_0 = 1.24 $ and $ n_0 = 0 $. 
		Dimensionful quantities are given in units of the pseudocritical temperature $ T_{pc} $ (cf.  Figure~\ref{fig:order_parameters}).
	}
	\label{fig:Evolution_dispersion_width}
\end{figure*}

The spectral functions can be used to deduce the dispersion relation and lifetimes of the corresponding quasiparticle species. Following \cite{aarts2001nonequilibrium} we assume for the moment that the spectral function decays exponentially and can be approximated as $ \rho(t, t', |\mathbf{p}|)  = e^{-\gamma_\mathbf{p}|t-t'|} \omega^{-1}_\mathbf{p} \sin [\omega_\mathbf{p}(t-t')]$ with a dispersion $  \omega_\mathbf{p}$ and a damping rate $  \gamma_\mathbf{p}$, which are both allowed to be $ \tau $ dependent. The corresponding Wigner transform is given by $ \rho(\tau,\omega, |\mathbf{p}|) = \rho_{\text{BW}}(\tau, \omega, |\mathbf{p}|) + \delta \rho (\tau, \omega, |\mathbf{p}|)$ where $\rho_{\text{BW}}  $ denotes the relativistic Breit-Wigner function
\begin{align}
\hspace{-.1cm}
\rho_{\text{BW}}(\tau, \omega, |\mathbf{p}|) 
= 
\dfrac{2\omega \Gamma(\tau, |\mathbf{p}|)}{\big[\omega^2 - \omega^2(\tau, |\mathbf{p}| )\big]^2 + \omega^2 \Gamma^2(\tau, |\mathbf{p}| )}
\,,
\label{eq:rel_BW}
\end{align}
which describes a peak with width $\Gamma(\tau, |\mathbf{p}|) = 2\gamma_\mathbf{p}(\tau) $
at position $\omega = \omega(\tau, |\mathbf{p}|) $. The term $\delta \rho  \sim \exp(-2 \tau \gamma_\mathbf{p})$  describes boundary effects due to the finite integration range in \eqref{eq:Wigner_transformation}. Since $ \delta \rho $ decreases exponentially with $ \tau \gamma_\mathbf{p}$, this term is negligible for sufficiently large damping ratios and/or sufficiently late times \cite{aarts2001nonequilibrium}. Otherwise, the frequency space spectral function suffers under severe noise coming from boundary effects. For all times shown in this work, we find that boundary effects are irrelevant.
 
We observe that peak shapes of the bosonic spectral functions can be well approximated by the Breit-Wigner function \eqref{eq:rel_BW}. At some given time $ \tau $, performing Breit-Wigner fits of the spectral function at all momenta $ |\mathbf{p} |$ yields the dispersion relation $ \omega(\tau, |\mathbf{p}|) $ and the momentum dependent width $ \Gamma(\tau, |\mathbf{p}|) $. 
For initial states with high energy densities, such as considered in this section, the spectral and statistical functions exhibit quasiparticle peak structures already at early times (see Figure~\ref{fig:Spectra_evolution_highT_pion}).
Consequently, it is possible to fit a Breit-Wigner function to the spectral functions at any stage such that the time evolution of the dispersion relation $ \omega_i(\tau, |\mathbf{p}|) $ and momentum dependent width $\Gamma_i(\tau, |\mathbf{p}|) $ for $ i = \sigma, \pi $ can be mapped out. 

In the left plot of Figure~\ref{fig:Evolution_dispersion_width} we show the dispersion relation of the pion at different times $ \tau $ encoded in the color scheme. A fit of $ \omega(\tau, |\mathbf{p}|)$ to the relativistic dispersion relation $ Z \sqrt{|\mathbf{p}|^2 + m^2} $ at various times $ \tau $ yields the quasiparticle masses $ m(\tau)$, which are shown in the inset. In the following, the stationary late-time value is denoted as $ m$.  
We note that the mass corresponds to the dispersion relation in the limit of vanishing momentum, i.e. $ m = \omega(\tau, |\mathbf{p}| \rightarrow 0) $. The right plot of Figure~\ref{fig:Evolution_dispersion_width} displays the momentum dependent width of the pion extracted from the Breit-Wigner fits. We find a plateau in the IR and a maximum in the UV. 
In analogy to the dispersion, where the quasiparticle mass describes the zero-momentum limit, we can extract the asymptotic value of the width in the limit of vanishing momentum, $ \Gamma =  \Gamma(\tau, |\mathbf{p}| \rightarrow 0) $. Since $ \Gamma $ corresponds to the width of the spectral function that is peaked at the quasiparticle mass, it can be viewed as the width of the quasiparticle. 
As the right plot in Figure~\ref{fig:Evolution_dispersion_width} indicates, $ \Gamma$ is increasing with time.

\begin{figure*}[t]
	\centering
	\includegraphics[width=1.\textwidth]{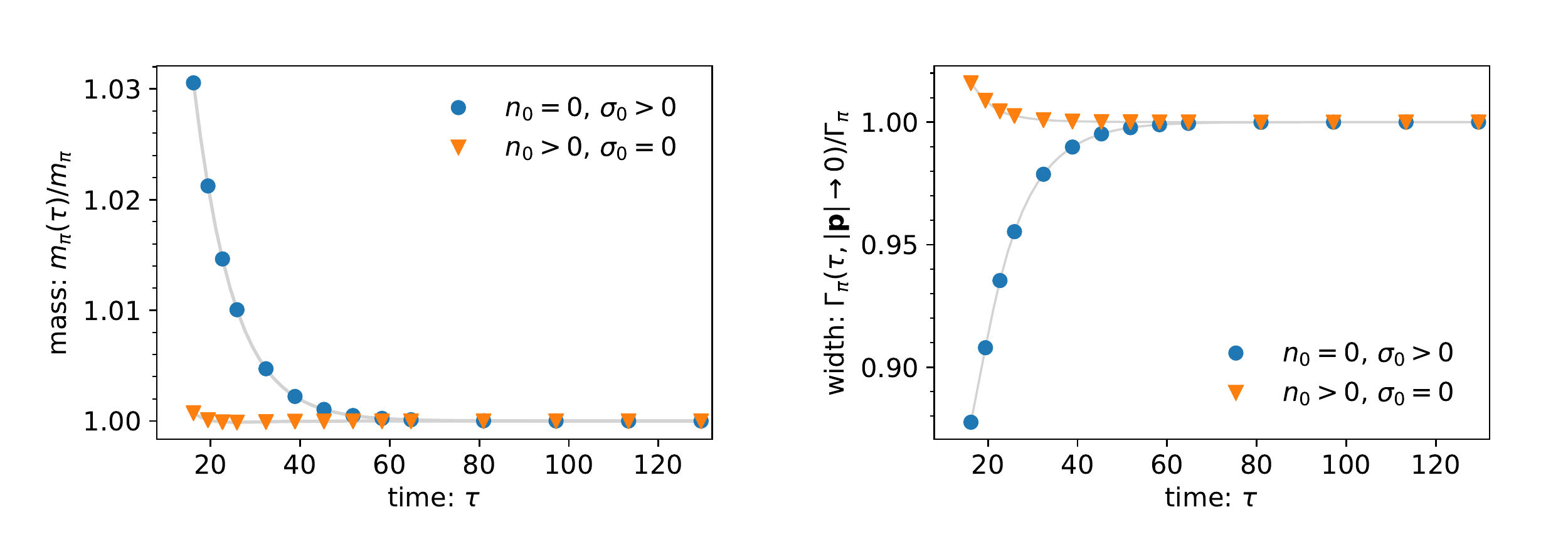}	
	\caption{
		Time evolution of the pion mass and the pion width in the limit $ |\mathbf{p}| \rightarrow 0 $. Results are shown for field dominated initial conditions with $ \sigma_0 = 1.36 $ and $ n_0 = 0 $ (blue dots) as well as fluctuation dominated initial conditions with $ \sigma_0 = 0 $ and $ n_0 = 0.8 $ (orange triangles).
		Dimensionful quantities are given in units of the pseudocritical temperature $ T_{pc} $ (cf.  Figure~\ref{fig:order_parameters}).
	}
	\label{fig:Evolution_mass_width}
\end{figure*}

We can now work out the differences observed in Figure~\ref{fig:Spectra_evolution_highT_pion} in a quantitative fashion. There is an apparent difference in the approach of the late-time values of the mass $ m_\pi$ and the width $ \Gamma_\pi  $ when comparing the time evolution starting from the two different initial conditions. The results are shown in Figure~\ref{fig:Evolution_mass_width}, where again only the pion data are shown because the sigma meson behaves accordingly.

For field dominated initial conditions the effective mass of the pion meson decreases during the time evolution, whereas for fluctuation dominated initial conditions it grows, albeit only slightly. This is in accordance to the previous observation of the shifting peak position for field dominated initial conditions.
It is important to note that the mass of the quasiparticles is not contained in the initial state, since $ m_\mathrm{init} $ is much smaller than the particle masses of the thermal state, but generated dynamically during the time evolution. The quasiparticle masses build up from the fluctuations contained in the self-energies. Since the nonzero initial field value leads to large bosonic self-energy contributions in the beginning of the time evolution, at early times the masses are larger than in the case of vanishing initial field.

The time dependence of the spectral width shown in the right plot of Figure~\ref{fig:Evolution_mass_width} can be understood in terms of the sum rule \eqref{eq:sum_rules} according to which the bosonic spectral functions are normalized. 
Due to the additional factor of $ \omega $ in the integrand, which arises from the time derivative on one of the fields in the boson commutation relation, a larger mass automatically implies smaller widths. Consequently, the behavior of mass and width in the time evolution must be converse to each other. 

After discussing the dynamics of the meson spectral and statistical functions at high initial energy densities, we now turn to the quark sector.
After decomposing the Dirac structure of fermionic two-point functions and imposing symmetries, we are dealing with four components for the quark spectral and statistical functions, the scalar, vector-zero, vector and tensor components as introduced in \eqref{eq:fermion_components}. Of these four components, the vector-zero component contains information about which states can be occupied \cite{tuprints2209, Kitazawa:2006zi}. Since it is normalized to unity according to the sum rule \eqref{eq:sum_rules}, the vector-zero component quark spectral function can be interpreted as the density of states for the quarks. 

We note that in a chiral symmetric theory with vanishing fermion bare mass one finds $ \rho_S  = \rho_P = \rho_T^{\mu \nu } = 0$ since only components in \eqref{eq:Lorentz_components} that anticommute with $ \gamma_5 $ are allowed. Here, we consider a setup where chiral symmetry restoration takes place. For initial conditions with high energy densities and the corresponding final states in the high-temperature chiral symmetric regime, the quark dynamics can be studied in terms of the vector-zero and vector components. 

\begin{figure*}[t]
	\centering
	\includegraphics[width=1.\textwidth]{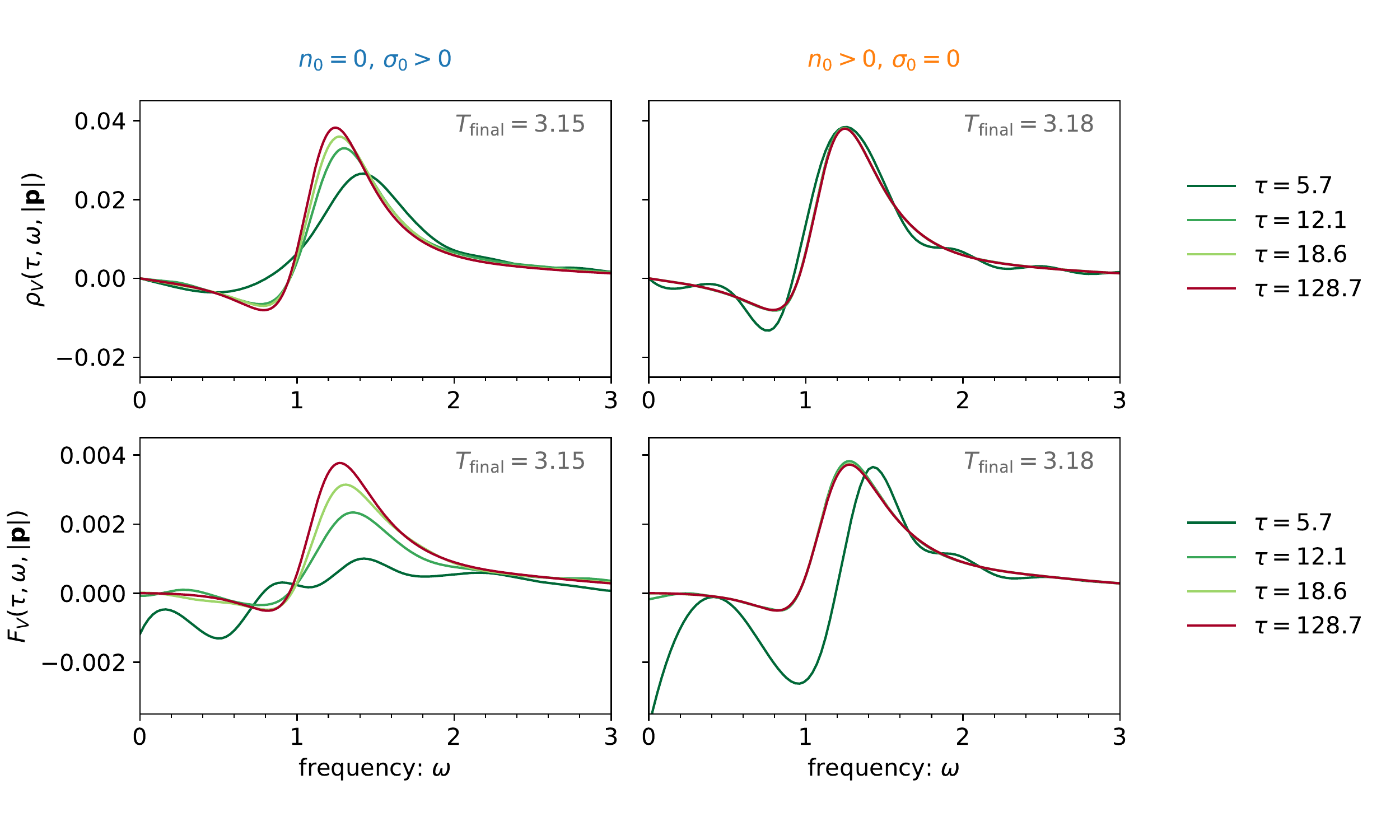}	
	\caption{
		Time evolution of the vector component quark spectral and statistical functions shown for the same initial conditions as in Figure~\ref{fig:Spectra_evolution_highT_pion} at momentum $ |\mathbf{p}| = 0.016 $. 
		Dimensionful quantities are given in units of the pseudocritical temperature $ T_{pc} $ (cf.  Figure~\ref{fig:order_parameters}).
	}
	\label{fig:Spectra_evolution_highT_V}
\end{figure*}

As was shown in Figure~\ref{fig:Spectra_evolution_summary}, for high energy densities there is not much dynamics taking place in the excitation spectrum of the quarks. More insight can be gained by looking at the vector component which is presented in Figure~\ref{fig:Spectra_evolution_highT_V} for the same field or fluctuation dominated initial conditions as discussed before for the bosons. 
The interesting case is again the evolution starting from field dominated initial conditions.  The corresponding vector spectral function (upper left plot) shows that the peak position moves toward smaller frequencies, just as in the bosonic case. It indicates that the energy of both meson and quark quasiparticles decreases during the time evolution. However, it is important to note that---in contrast to the mesons---the amplitude of the fermion statistical function increases during the time evolution. As discussed before, the nonzero initial field leads to strong fluctuations and hence occupancies in the bosonic sector. It takes time for these fluctuations to be transferred to the fermionic sector, which is why we observe that the fermion occupation grows slowly during the time evolution. 

For the fluctuation dominated initial conditions, we again observe that the spectral and statistical functions approach their late-time behavior very quickly. We conclude that the available states and their occupation quickly approach their thermal final state if energy is provided in terms of particles rather than the field in the initial state. 

\subsubsection{Intermediate energy densities}
\label{sec:intermediate_energy_densities}

From Figure~\ref{fig:Spectra_evolution_summary} we can see that the most interesting dynamics is taking place for systems thermalizing in the crossover region. Thus, we aim to study the evolution of the vector-zero quark spectral function for two simulations thermalizing in the cross-over region. 

Again we compare two simulations employing field or fluctuation dominated initial conditions, respectively, but in this case we are able to probe initial conditions that lead to the same late-time state. 
When comparing the late-time field expectation value $ \overline{\sigma} $, the mass ratio $ m_\sigma / m_\pi  $, and the temperature $ T $ of the final state of these two simulations, we find that the respective quantities differ by less than $ \SI{0.5}{\percent} $. Also, the shapes of the spectral and statistical functions in frequency space are the same for both bosonic as well as fermionic components. Quantitatively, we find that $ | \rho_1 - \rho_2 | /\max{(\rho_1)} $ is smaller than $ \mathcal{O}(10^{-2}) $ for all frequencies $ \omega $ and momenta $ |\mathbf{p}| $, where the indices 1 and 2 denote the two simulations compared and $ \max{(\rho)} $ the maximal amplitude of $ \rho $. Larger deviations are observed for the vector-zero component statistical function and for the tensor component spectral and statistical functions, where the amplitudes are of order $ \mathcal{O}(10^{-7}) $ such that numerical inaccuracies come into play. In conclusion, we consider the late-time state of the two simulations to be the same thermal state, universal in the sense that the dependence on the initial conditions is lost. It is characterized solely by a temperature of $ T = 1.04 $, a mass ratio of $ m_\sigma / m_\pi = 1.46 $ and a field expectation value of $ \overline{\sigma} = 0.33 $. As we will see later, this corresponds to a state in the crossover region.

\begin{figure*}[t]
	\centering
	\includegraphics[width=1.\textwidth]{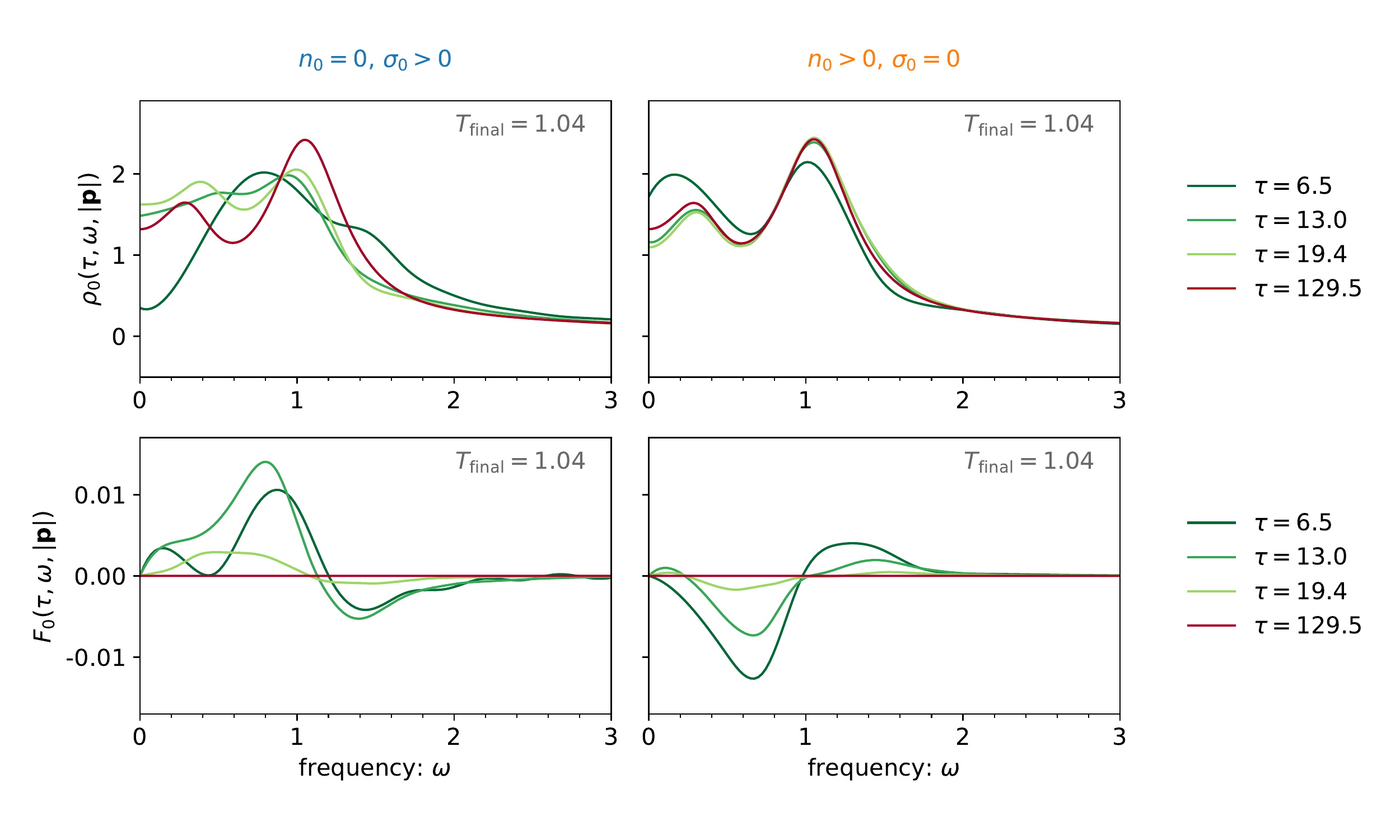}	
	\caption{
		Time evolution of the vector-zero component quark spectral and statistical functions shown for two different initial conditions at momentum $ |\mathbf{p}| = 0.012 $. The left column shows field dominated initial conditions with $ \sigma_0 = 0.98 $, the right column fluctuation dominated initial conditions with $ n_0 = 0.11$. Both lead to the same late-time state with $ T = 1.04 $. 
		Dimensionful quantities are given in units of the pseudocritical temperature $ T_{pc} $ (cf.  Figure~\ref{fig:order_parameters}).
	}
	\label{fig:Spectra_evolution_crossover_V0}
\end{figure*}

The regime of intermediate energy densities distinguishes itself from high and low energy density initial conditions by showing a double-peak structure in the quark spectral functions. Our findings in a nonperturbative real-time setting corroborate previous observations of such double peak structures with perturbative computations or spectral reconstructions reported, e.g., in \cite{Kitazawa:2005mp,Kitazawa:2006zi,Kitazawa:2007ep,Karsch:2009tp,Mueller:2010ah,Qin:2010pc,Qin:2013ufa,Fischer:2017kbq}.

First, consider the vector-zero component describing the excitation spectrum of the quarks. In Figure~\ref{fig:Spectra_evolution_crossover_V0} we show the time evolution of both spectral and statistical functions. As before, for fluctuation dominated initial conditions (right column) the system quickly approaches the shape of the late-time two-point functions. 
However, in the case of field dominated initial conditions, the double-peak structure of the spectral function only emerges at later times. At early times, the spectral function reveals a single broad structure. 

We further point out that the statistical function $ F_0 $ decays to zero during the time evolution, implying that the fermion occupation is not contained in the vector-zero component but in other components. This agrees well with the effective quasiparticle number that has been employed previously \cite{tuprints2209, Berges:2010zv}, 
\begin{align}
\hspace{-.1cm}
n_\psi(t, |\mathbf{p}|) = \dfrac{1}{2}
- 
\dfrac{|\mathbf{p}| F_V(t, t, |\mathbf{p}|) + M_\psi(t)F_S(t,t,|\mathbf{p}|)}{\sqrt{|\mathbf{p}|^2 + M_\psi^2(t)}}\,,
\end{align} 
with effective mass $ M_\psi(t) = m_\psi + h \sigma(t) $. This definition of an effective particle number only provides a good description of the quark content in the system if the occupations in the vector-zero and tensor component are negligible. In our computations, we find that $ F_0 $ and $ F_T $ are of the order $ \mathcal{O}(10^{-7}) $ and hence irrelevant for the quark particle number. 

\begin{figure*}[t]
	\centering
	\includegraphics[width=1.\textwidth]{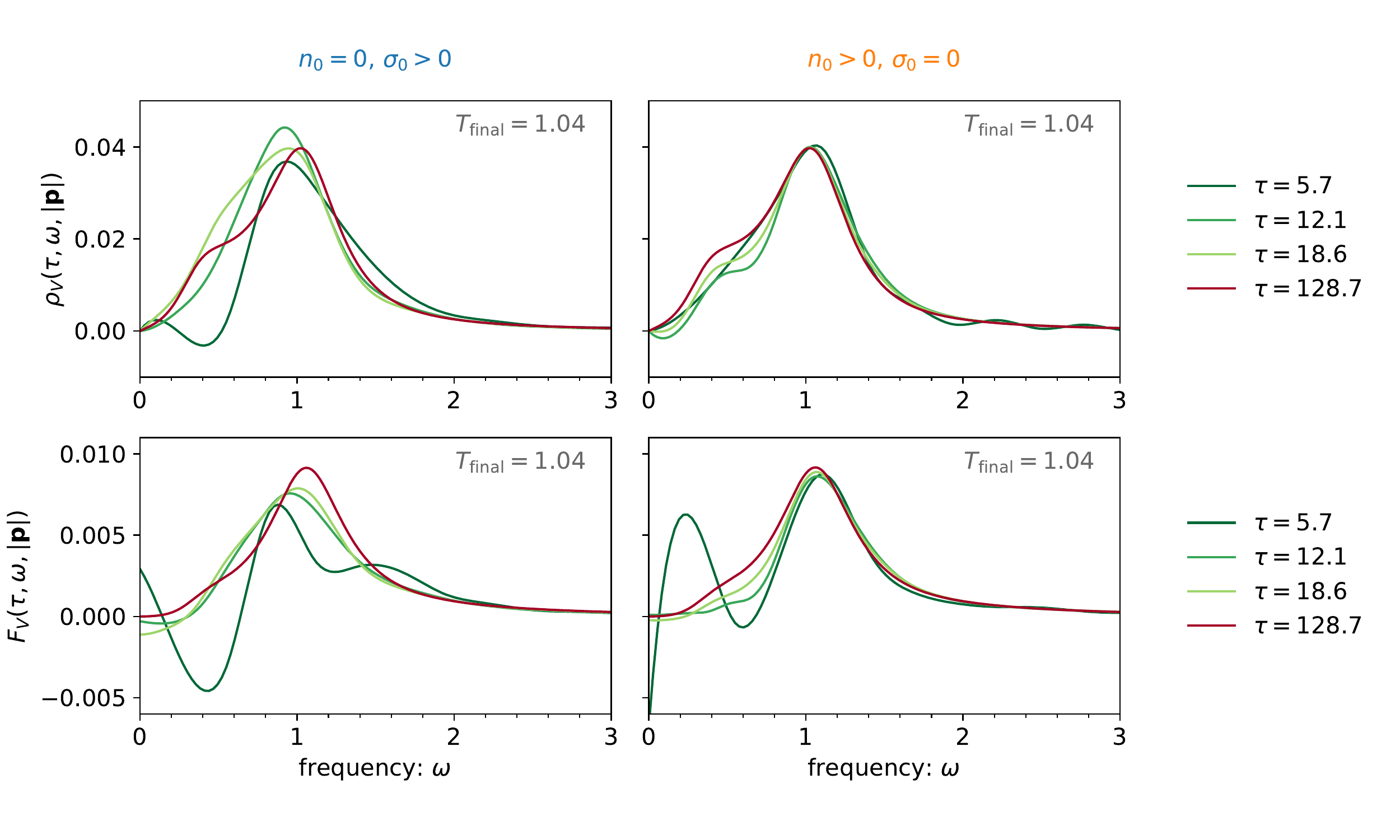}	
	\caption{
Time evolution of the vector component quark spectral and statistical functions shown for two different initial conditions at momentum $ |\mathbf{p}| = 0.012 $. The left column shows field dominated initial conditions with $ \sigma_0 = 0.98 $, the right column fluctuation dominated initial conditions with $ n_0 = 0.11$. Both lead to the same late-time state with $ T = 1.04 $. 
		Dimensionful quantities are given in units of the pseudocritical temperature $ T_{pc} $
		}
	\label{fig:Spectra_evolution_crossover_V}
\end{figure*}

In order to study the particle content, we take into account the vector component which is shown in Figure~\ref{fig:Spectra_evolution_crossover_V}. We can see that the double-peak structure observed in the vector-zero component is also visible in the vector component, in particular in both spectral and statistical functions. From this, we learn that the additional light degrees of freedom, provided in the low-frequency peak of the quark spectral density, is actually occupied in terms of the vector component quark statistical function. Hence, for states thermalizing in the crossover temperature regime, there is an additional light mode with finite occupation in the quark sector available to participate in the dynamics.

We further observe that for fixed momentum $ |\mathbf{p}| $ the energy of the light mode increases with rising temperature. At sufficiently high temperatures this additional mode reaches energies comparable with the main quasiparticle mode such that the two peaks merge into the single peak persistent in the high-temperature regime. 

We conclude this section with a comment on the dynamics found for initial states with low energy-densities. In contrast to the cases of intermediate and high energy densities, we find well-defined quasiparticle peaks for both quarks and mesons. 
The smaller energy density leads to lower thermalization temperatures and a stronger chiral symmetry breaking, reflected by a mass difference between the $ \sigma $ and $ \pi $ mesons. After discussing the nonequilibrium time evolution of the spectral functions, we now turn to the equilibrium properties.

\subsection{Late-time thermal limit}
\label{sec:equilibrium_spectra}

In this section we discuss the spectral functions of quarks and mesons in the state of quantum thermal equilibrium according to the definition introduced in Section~\ref{sec:thermal_equilibrium}. The properties of spectral functions at different temperatures reflect the crossover transition of the quark-meson model from the chiral broken to a chiral symmetric phase.  We find that the shapes of the final states are universal in the sense that they only depend on the temperature and not on the details of the initial state.

\subsubsection{Mesons}
\label{sec:equilibrium_spectra_mesons}

\begin{figure*}[t]
	\centering
	\includegraphics[width=1.\textwidth]{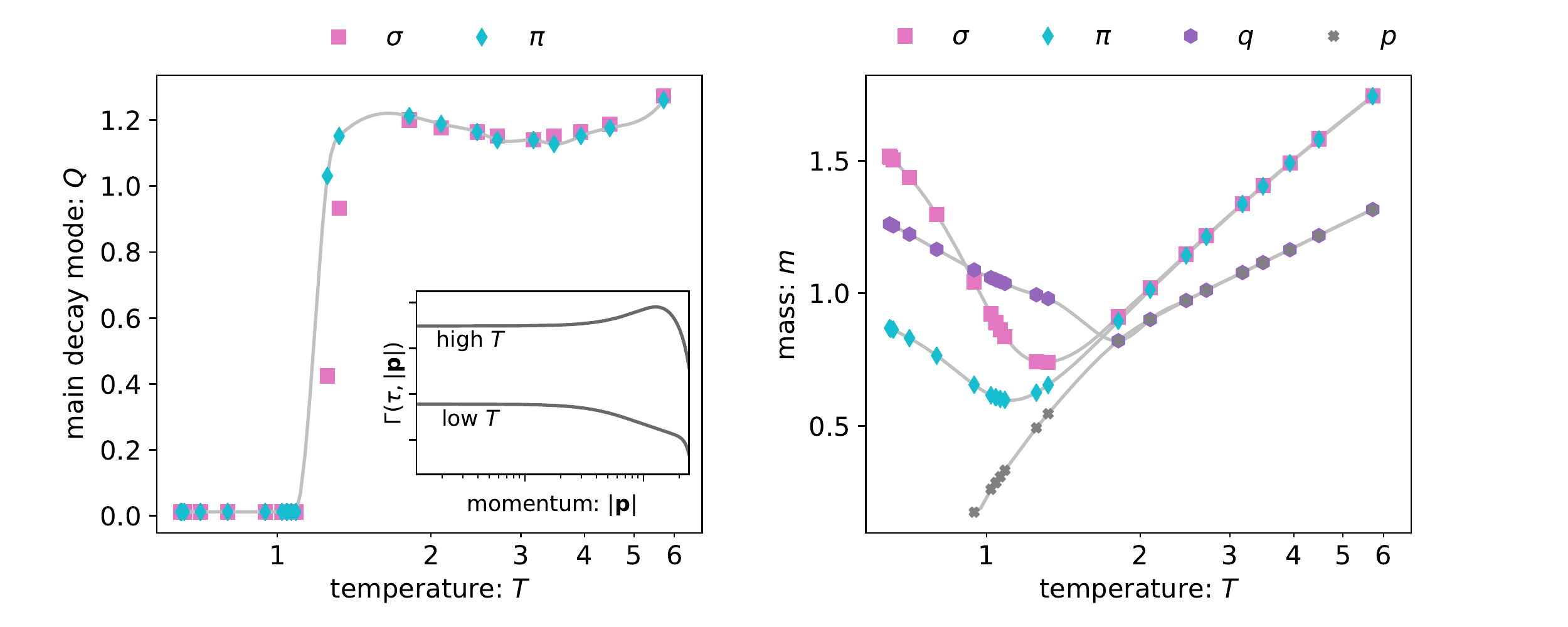}	
	\caption{
		Left: Temperature dependence of the characteristic decay momentum $ Q $ shown for the $ \sigma $ and $ \pi $ mesons. The inset shows examples for the momentum dependent width at high and low temperatures. $ Q $ corresponds to the momentum at which the width $ \Gamma(\tau, |\mathbf{p}|) $ is maximal. \\
		Right:
		Temperature dependence of quasiparticle masses. 
		Restoration of chiral symmetry is reflected in identical masses of the $ \sigma $ and $ \pi $ mesons at high temperatures. 
		The quark $q$ quasiparticle mass is obtained from the dominant peak of the vector-zero component quark spectral function. We also plot the ``plasmino" branch $p$ obtained from the quark spectral function.
		In both plots gray lines show cubic spline interpolations of the data points. Dimensionful quantities are given in units of the pseudocritical temperature $ T_{pc} $ (cf. Figure~\ref{fig:order_parameters}).
		}
	\label{fig:equilibrium_boson_width_mass}
\end{figure*}

Information about the different phases of the model can be obtained from the temperature dependence of the late-time thermal spectral functions of the mesons. 
We find that the shape of the bosonic spectral functions is described by a Breit-Wigner function for all considered temperatures. Thereby, the width and the position of the Breit-Wigner peak only depend on the temperature but not on the initial conditions chosen. 

As discussed in Section~\ref{sec:high_energy_densities}, the momentum dependent width and the dispersion relation are obtained by applying Breit-Wigner fits to the spectral functions. 
Although the Breit-Wigner function \eqref{eq:rel_BW} has two parameters, the width $ \Gamma(\tau, |\mathbf{p}|) $ and the peak position given by $ \omega(\tau, |\mathbf{p}|) $, there is only one free parameter since the normalization condition given by the sum rule \eqref{eq:sum_rules} must be satisfied. 

In the right plot of  Figure~\ref{fig:Evolution_dispersion_width} we already saw that there is a characteristic momentum mode $ |\mathbf{p}| $ at which the momentum dependent width becomes maximal. This corresponds to the momentum at which the decay is strongest and can be considered as the \textit{main decay mode}, in the following denoted by $ Q $. 
In the left plot of Figure~\ref{fig:equilibrium_boson_width_mass} we show the main decay mode $ Q $ as a function of temperature for both meson species. 
At low temperatures, the strongest decays are found in the IR, whereas at high temperatures the strongest decays occur in the UV. There is an abrupt change at some critical temperature, above which $ Q>0 $ meaning that the momentum dependent width has a maximum at a nonzero momentum, as shown by the upper line in the inset. Comparing the momentum dependent width at low $ T $ and high $ T $, we can see that the transition from the chiral broken to the chiral symmetric phase is characterized by new decay modes in the UV. Thereby, the main decay mode is suddenly shifted from the IR to the UV. 

Another prominent signature for the crossover transition is provided by the quasiparticle masses of the $ \sigma $ and $ \pi $ mesons. 
The two meson species are distinguishable in the chiral broken phase, where they have different masses, while they become identical in the chiral symmetric phase. 
When plotting the meson masses as a function of temperature, as shown in the right plot of Figure~\ref{fig:equilibrium_boson_width_mass}, we can nicely visualize the restoration of chiral symmetry, manifest in the quasiparticle masses of $ \sigma $ and $ \pi $ becoming identical (pink and cyan data points). 
We observe a softening of the masses at intermediate temperatures, i.e. the quasiparticle masses are minimal in the temperature region where the crossover phase transition occurs. 
Decreasing masses indicate growing correlation lengths. 
In the limit of a second order phase transition, which is characterized by diverging correlation lengths, the masses would vanish at the transition point. 
In the high-temperature range, masses grow with rising temperatures. 
This reflects that the quasiparticle masses can be considered as \textit{thermal masses} in the sense that they contain self-energy contributions and are generated by quantum fluctuations, which increase with temperature. 

We further note that one could also study the temperature dependence of the width $ \Gamma = \Gamma(\tau, |\mathbf{p}|\rightarrow 0) $ instead of $ m = \omega(\tau, |\mathbf{p}| \rightarrow 0) $. However, the information is equivalent due to the normalization of the spectral functions, as pointed out above. Consequently, the behavior of $ \Gamma $ is converse to the behavior of $ m $ and not presented here explicitly. 
The width $ \Gamma $ is small at low temperatures, strongly grows toward intermediate temperatures where it reaches a maximum value in the crossover temperature regime, and then decays slowly when going to higher temperatures.

\subsubsection{Quarks}

\begin{figure*}[t]
	\centering
	\includegraphics[width=1.\textwidth]{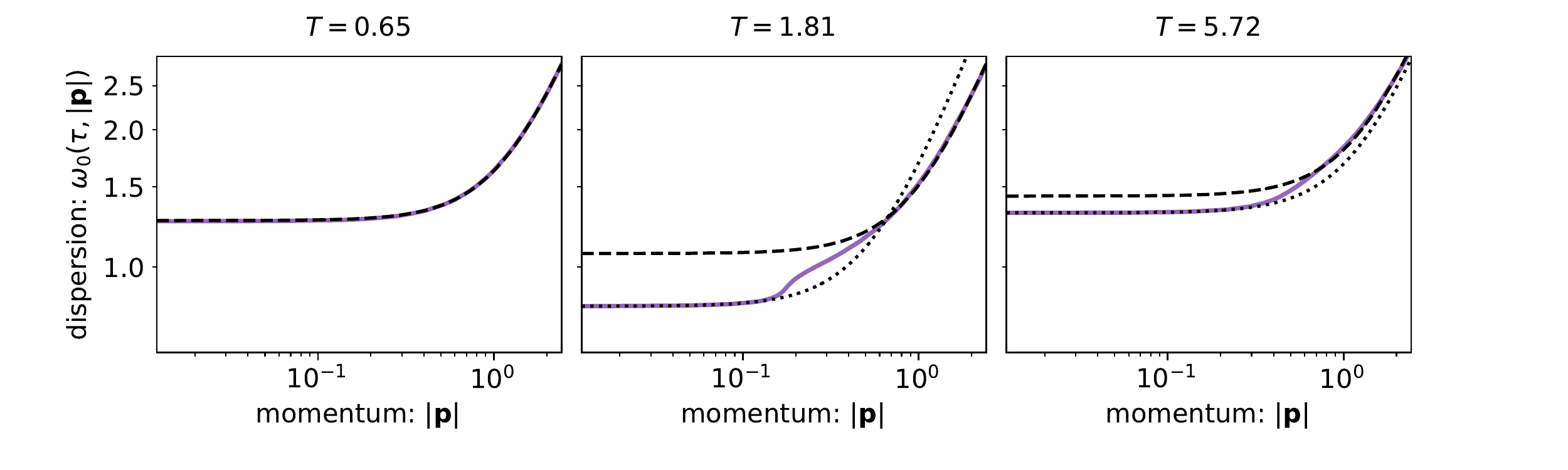}	
	\caption{
		The dispersion relation of the vector-zero quark spectral function shown for three different temperatures.
		At low temperature a fit to the relativistic dispersion relation $ Z\sqrt{|\mathbf{p}|^2 + m_q^2} $ is shown by the black dashed line. For higher temperatures the behavior at small and large momenta differs as the additional low-frequency peak and the main peak merge into one peak. 
		We perform separate fits at low and high momenta, shown by the dashed and dotted black lines. 
		Dimensionful quantities are given in units of the pseudocritical temperature $ T_{pc} $ (cf.  Figure~\ref{fig:order_parameters}).}
	\label{fig:dispersion_V0}
\end{figure*}

\begin{figure*}[t]
	\centering
	\includegraphics[width=1.\textwidth]{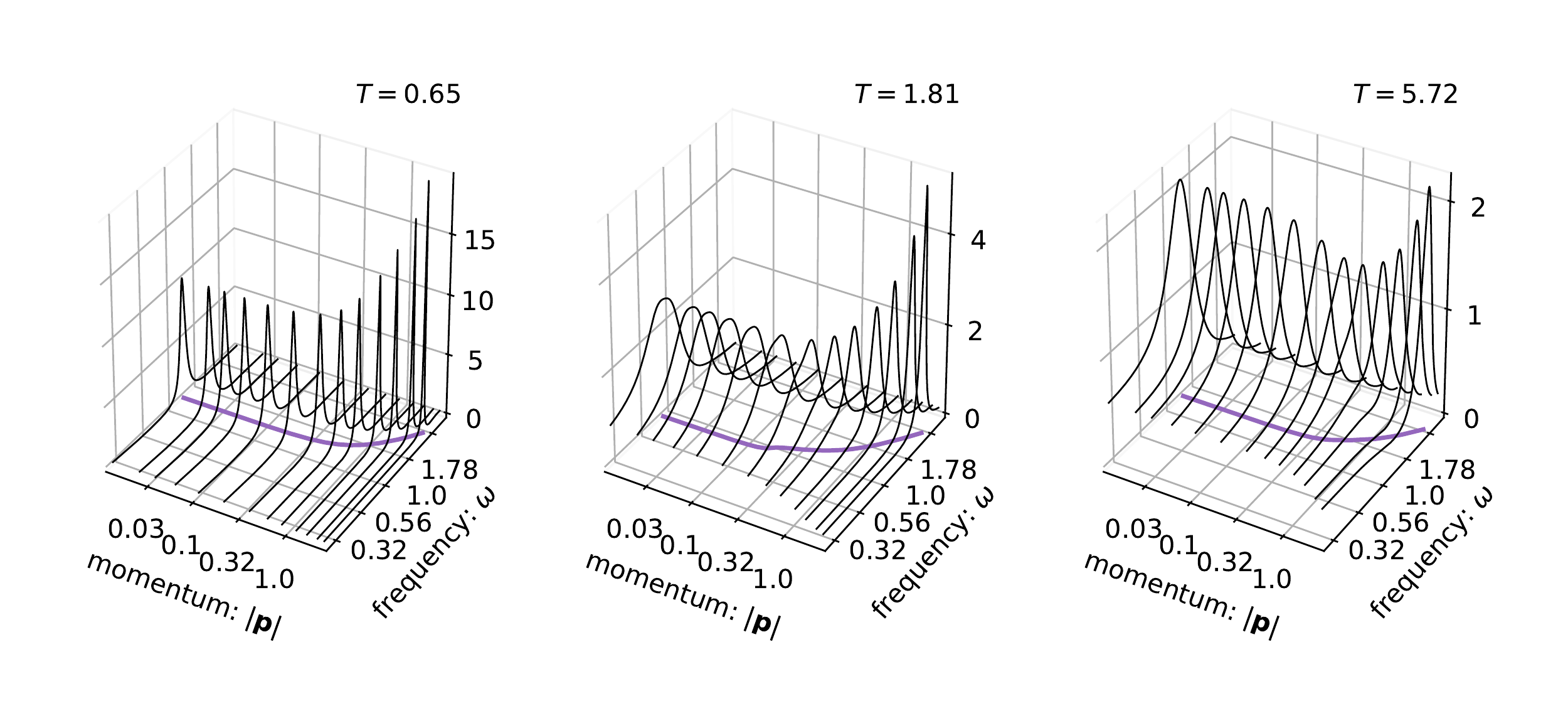}	
	\caption{
		The vector-zero quark spectral function as a function of frequency $ \omega $ shown for a range of spatial momenta $ |\mathbf{p}| $. The three plots correspond to the same three temperatures as in Figure~\ref{fig:dispersion_V0}. 
		The purple line indicates peak position of the spectral function in the $|\mathbf{p}|$-$ \omega $-plane and is therefore equivalent to the dispersion relation shown in Figure~\ref{fig:dispersion_V0}. 
		The spectral function reveals a narrow quasiparticle peak at low temperatures. As the temperature is increased the light mode interferes with the low-momentum spectral function, leading to a broad peak at small momenta. At high momenta, the quasiparticle peak remains narrow. 
		Dimensionful quantities are given in units of the pseudocritical temperature $ T_{pc} $ (cf.  Figure~\ref{fig:order_parameters}).	}
	\label{fig:Spectral_function_V0_3D}
\end{figure*}

Let us now consider the thermal spectral functions for the quark sector. Several aspects of the different components invite for discussion. Let us begin with a recap of the findings shown in the vector-zero component of the quark spectral function. As presented in Figure~\ref{fig:Spectra_evolution_summary} the spectral density has different shapes at low, intermediate and high temperatures. In particular, the intermediate temperature range of the crossover transition is characterized by a double-peak structure. 
The temperature dependence of the fermionic quasiparticle masses is depicted in Figure~\ref{fig:equilibrium_boson_width_mass}. The mass of the low-frequency mode (plasmino branch, denoted by $p$) grows continuously with rising $ T $ until it merges with the main peak (denoted by $q$), forming the wide quasiparticle peak found for initial states with large energy densities. For related studies with perturbative computations or spectral reconstructions see,  e.g.,~\cite{Kitazawa:2005mp,Kitazawa:2006zi,Kitazawa:2007ep,Karsch:2009tp,Mueller:2010ah,Qin:2010pc,Qin:2013ufa,Fischer:2017kbq}. Note also that this double-peak structure is only visible in the small momentum regime. It can be studied by considering the dispersion relation obtained from the vector-zero quark spectral function.

For temperatures below some critical temperature in the crossover regime, the vector-zero spectral function  
reveals the shape of a nonrelativistic Breit-Wigner function, also known as the Lorentz function, 
\begin{align}
\rho_{\text{L}}(\tau, \omega, |\mathbf{p}|) 
= 
\dfrac{A\  \Gamma(\tau, |\mathbf{p}|)}{\big[\omega - \omega(\tau, |\mathbf{p}| )\big]^2 + \Gamma^2(\tau, |\mathbf{p}| )}
\,,
\label{eq:Lorentz}
\end{align}
where $ A $ is a normalization constant, $ \Gamma(\tau, |\mathbf{p}|) $ the width and $\omega(\tau, |\mathbf{p}| )  $ the dispersion. When temperature is increased, the vector-zero quark spectral function ceases to be described in terms of \eqref{eq:Lorentz} as the low-frequency mode arises and grows in amplitude. 
Due to appearance of the additional peak, it is not possible to perform a Lorentz fit at all temperatures. 
As a consequence, we choose to compute the dispersion relation of the quarks by determining the peak position of the main peak of $ \rho_0 $. The obtained dispersion relation is shown for three temperatures in Figure~\ref{fig:dispersion_V0}. 

For low temperatures, where no additional peak is present, the quark dispersion is well described by a relativistic dispersion relation; see left plot of Figure~\ref{fig:dispersion_V0}.
When going to intermediate temperatures, the additional light mode leads to a double-peak structure. As long as the two peaks are distinguishable, one can determine the dispersion relation of the main peak, which yields the same shape as in the low-temperature regime. 
However, when the main peak and the side peak merge into a single peak, the dispersion relation obtained from the overlap of the two peaks has a dispersion relation of the form shown by the middle plot of Figure~\ref{fig:dispersion_V0}. There is a clearly visible dip in the dispersion, showing that for small momenta the peak position is determined by the light mode, while for large momenta the peak position is determined by the main peak. 
We can fit the low-momentum and high-momentum areas separately to a relativistic dispersion relation, as shown by the dashed and dotted lines in Figure~\ref{fig:dispersion_V0}. 
When considering higher temperatures, the position of the dip moves toward larger frequencies and is not visible by eye anymore. However, we find that the dispersion relation cannot be described by the relativistic dispersion relation $ Z \sqrt{|\mathbf{p}|^2 + m^2} $ over the whole momentum range but still distinguishes between high-momentum and low-momentum regimes. 
We conclude that the single peak of $ \rho_0 $ at large temperatures is still the result of an overlap of a small low-frequency peak with the main peak. 
More insight is gained by considering the momentum dependence of the corresponding spectral functions, which is shown in Figure~\ref{fig:Spectral_function_V0_3D}. The spectral function $ \rho_0(\tau, \omega, |\mathbf{p}|) $ is shown at some late time $ \tau $ where the system has approached thermal equilibrium. The peak position of the spectral functions corresponds to the dispersion relations shown in Figure~\ref{fig:dispersion_V0}. At low temperatures, we find a single narrow quasiparticle peak. For higher temperatures, however, an additional light mode interferes with the main peak. At intermediate temperatures, where a softening of the mass occurs, the light mode and the main peak have comparable frequencies in the infrared. The superposition of the main peak and the light mode leads to a broad peak at small momenta, whereas the peak remains narrow at high momenta. 
As temperature increases, the light mode is only visible at higher momenta. An example is shown by the right plot in Figure~\ref{fig:Spectral_function_V0_3D}, where one can see a small enhancement of the spectral function at low frequencies for intermediate momenta. 
This observation indicates that the quark spectral function harbors additional degrees of freedom at high temperatures, as compared to the low-temperature regime. 

From the dispersion relation of the vector-zero quark spectral function, we determine the constituent quark mass by taking the asymptotic value at vanishing momentum, i.e. $ m_q = \omega_0(\tau,|\mathbf{p}| \rightarrow0 ) $. 
The constituent quark mass behaves analogously to the bosonic masses, i.e., a softening of the mass in the crossover temperature range occurs, see violet data points in the right plot of Figure~\ref{fig:equilibrium_boson_width_mass}. 
At low temperatures, the constituent quark mass lies between the $ \sigma $ and $ \pi $ masses, which is in qualitative agreement with the particle masses known at $ T=0 $. 
For temperature below $ T \simeq 2 $ we find that the pion is the lightest particle in the theory. This supports chiral perturbation theory as an effective theory for QCD where only pion degrees of freedom are considered. 
On the contrary, at high temperatures the constituent quark mass is smaller than the meson mass. As light modes are easier to excite, they dominate the dynamics in a system. Hence, our observation matches our idea that the chiral symmetric phase is dominated by quark degrees of freedom whereas the chiral broken phase is described by hadronic degrees of freedom, in particular by pions.

\begin{figure*}[t]
	\centering
	\includegraphics[width=1.\textwidth]{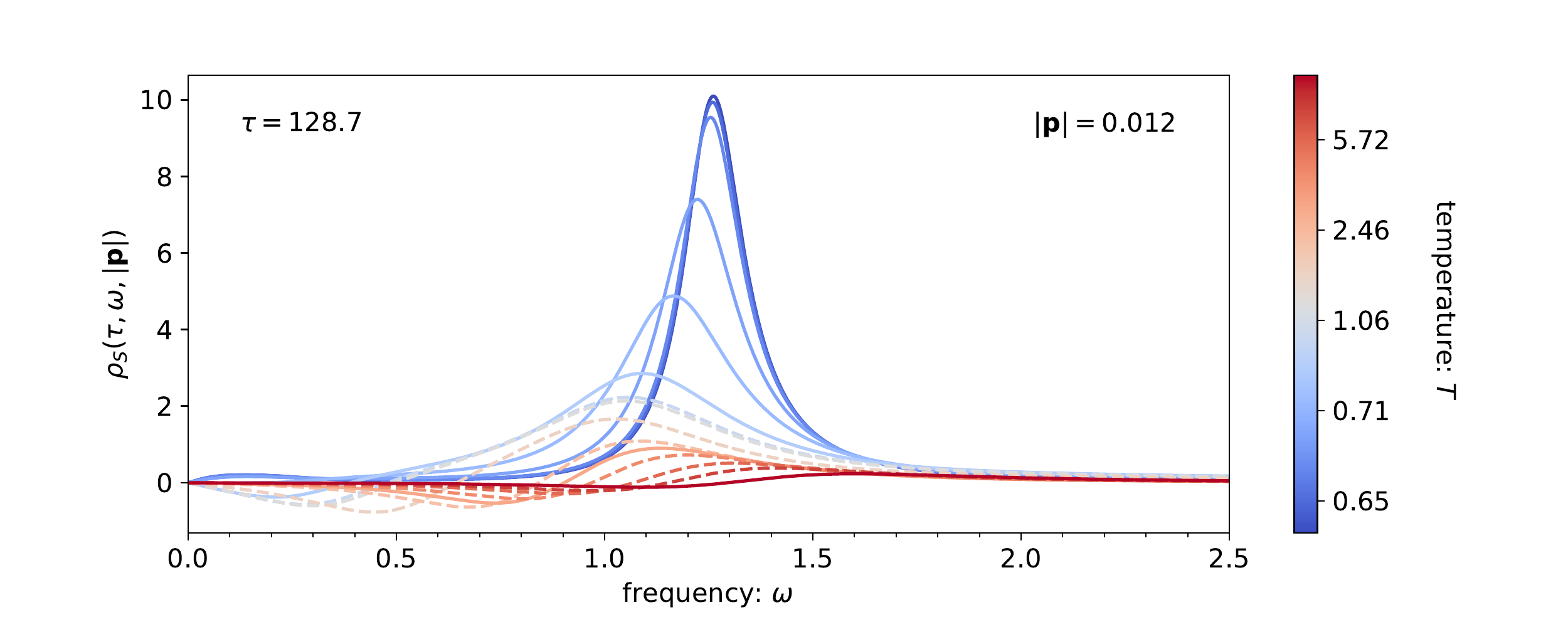}	
	\caption{
		The thermalized scalar component of the quark spectral function as a function of frequency shown for different temperatures. Dimensionful quantities are given in units of the pseudocritical temperature $ T_{pc} $ (cf.  Figure~\ref{fig:order_parameters}).
	}
	\label{fig:Spectra_thermal_S}
\end{figure*}

Finally, we shortly discuss the scalar component of the quark spectral function. 
In a chiral symmetric theory with vanishing fermion bare mass, the scalar component of the quark spectral function vanishes, i.e. $ \rho_S = 0 $. Although chiral symmetry is broken explicitly here, we expect the system to restore chiral symmetry at high temperatures, implying that the limit of a vanishing scalar component quark spectral function is approached. 
In Figure~\ref{fig:Spectra_thermal_S} we present the numerical results for a range of temperatures. The clear quasiparticle peak existing at low temperatures widens and flattens with rising temperature. The amplitude of the scalar component finally decays to zero, visualizing the predicted restoration of chiral symmetry in the course of the crossover transition. 
We further note that the peak position of the scalar component spectral function qualitatively shows the same behavior as the vector-zero component. The peak moves toward small frequencies at intermediate temperatures, corresponding to the softening of a mass, and is shifted toward higher frequencies at low and high temperatures. 

\section{The macroscopic field}
\label{sec:field}

In this section, we study the time evolution of the expectation value of the macroscopic field $\braket{ \sigma(t)} $, for the set of different initial conditions deployed also in the previous section. In addition, we study the model for different fermion bare masses in order to analyze the effects of spontaneous symmetry breaking in the model. 

\subsection{Nonequilibrium time evolution of the field}

\begin{figure*}[t]
	\centering
	\includegraphics[width=1.\textwidth]{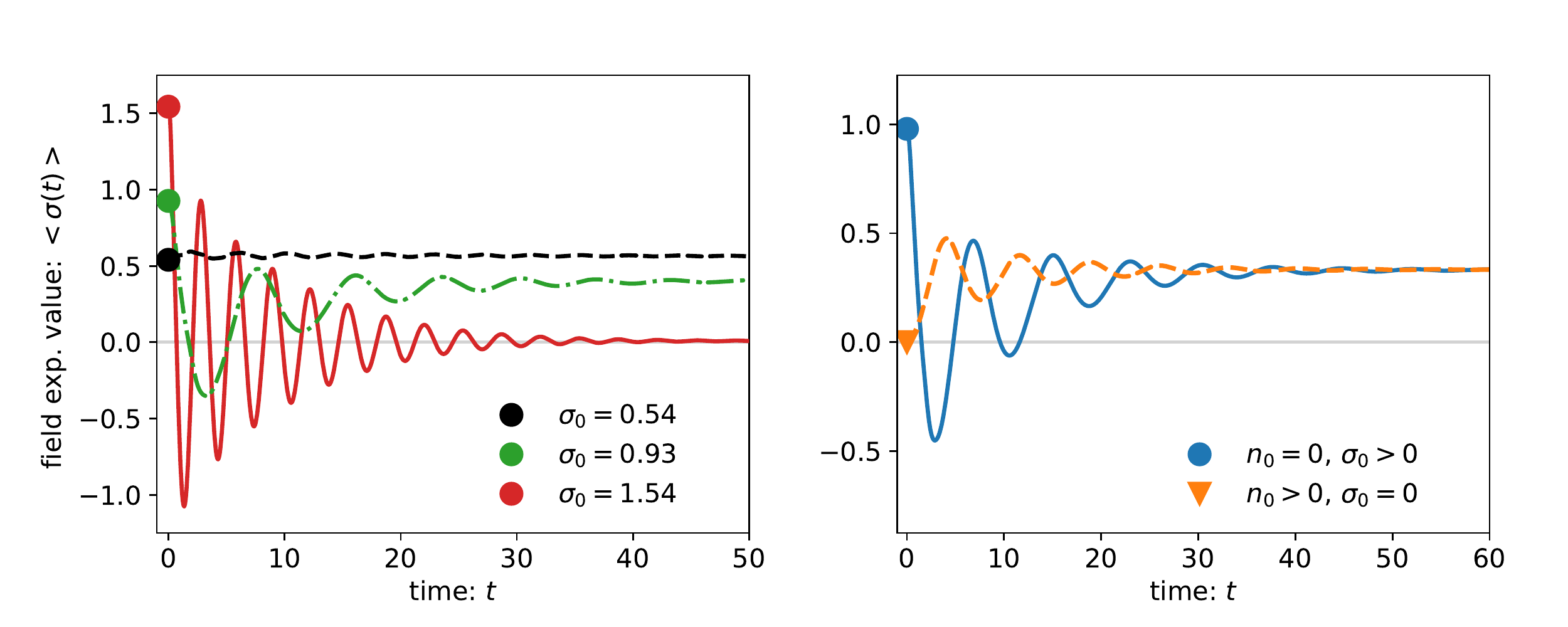}	
	\caption{
		Left: The time evolution of the field shown for field dominated initial conditions with different initial field values as indicated in the legend. 
		Right: The time evolution of the field shown for field dominated initial conditions with $ \sigma_0 = 1.98 $ (blue) and fluctuation dominated initial conditions with $ n_0 = 0.11 $ (orange). The same late-time field value $ \bar{\sigma} = 0.33 $ is approached for both initial states. 
		The gray line in both plots serves as a guide to the eye for $ \braket{\sigma(t)}= 0 $. 
		Dimensionful quantities are given in units of the pseudocritical temperature $ T_{pc} $ (cf.  Figure~\ref{fig:order_parameters}).
		}
	\label{fig:field_evolution_IC}
\end{figure*}

The classical potential of the sigma field is given by
\begin{align}
V(\sigma) = \dfrac{1}{2} m^2 \sigma^2 + \dfrac{\lambda}{4!N}\sigma^4\,,
\label{eq:classical_potential}
\end{align}
where the parameter choice of $ m^2 < 0 $ allows for spontaneous symmetry breaking. Thus, the potential has the shape of a double well with minima located at $ \sigma= \pm \sqrt{- 3!N \,m^2 / \lambda }  $. For the parameters employed in this work the minimum is located at $ \sigma \approx 0.04 $. 
The time evolution of a classical field in this potential is described by the classical equation of motion 
\begin{align}
\left[	
\partial_t^2 + m^2 + \dfrac{\lambda}{6N} \sigma^2(t) \right]  \sigma(t)
=	0\,,
\label{eq:classical_eom_field}
\end{align}
where spatial homogeneity and isotropy are assumed. If the initial field value, or the initial field derivative, is nonzero, the field rolls down a potential hill and oscillates until it equilibrates at the minimum of the potential. 
Here, we go beyond the classical theory and compute the nonequilibrium time evolution including additional quantum fluctuations. As discussed above, we employ an approximation that includes quantum corrections at NLO in $ 1/N $ and $ g $. The quantum corrections lead to an effective potential and additional terms in the field equation \eqref{eq:classical_eom_field}. 
The full evolution equation at the given approximation can be found in Appendix~\ref{app:eom}. 

Depending on the initial conditions the time evolution of the field shows different properties. 
Let us first consider field dominated initial conditions, where the initial field is set to a finite value $ \sigma_0 $. The time evolution for the expectation value of the field $ \braket{\sigma(t)} $ is shown for different $ \sigma_0 $ in the left plot of Figure~\ref{fig:field_evolution_IC}. One can see that the field oscillates and eventually reaches a stationary value. 
In contrast to the classical theory, where the field always reaches the same equilibrium value given by the position of the potential minimum, the field reaches different late-time values. 
The reason is that the field itself generates quantum fluctuations as it rolls down a potential hill. These dynamically emerging fluctuations again influence the effective potential in which the nonequilibrium time evolution takes place. As the initial field value effects the amount of quantum fluctuations in the system and hence the shape of effective potential, different values of $ \sigma_0 $ lead to different late-time values for $ \braket{\sigma(t)} $. Before we come to a more detailed discussion of the plots in Figure~\ref{fig:field_evolution_IC}, we provide some intuition for the influence of fluctuations on the effective potential. 

Quantum fluctuations can be represented as loop corrections of the effective action. The effective potential is obtained when evaluating this effective action at a constant field. 
For a nonvanishing fermion bare mass, i.e. $ m_\psi \neq 0 $, the chiral symmetry breaking tilts the effective potential toward negative values. Thereby, larger $ m_\psi $ cause stronger tilts. On the other hand, bosonic fluctuations provide positive contributions, pushing the potential toward a symmetric shape. Together, this leads to a tilted Mexican hat potential with a minimum at some finite field expectation value. The position of the minimum of the effective action can easily be much larger than the position of the minimum of the classical potential.

The influence of these quantum corrections to the effective potential can be visualized by looking at the energy density of the system, which we compute from the energy-momentum tensor. We distinguish classical, bosonic and fermionic contributions to the energy density, with the relevant expressions presented in Appendix~\ref{sec:EMT}. Quantum fluctuations are taken into account for the fermionic and bosonic parts of the energy density. Hence, the energy density reflects the amount of fluctuations in the system. 

\begin{figure*}[t]
	\centering
	\includegraphics[width=.5\textwidth]{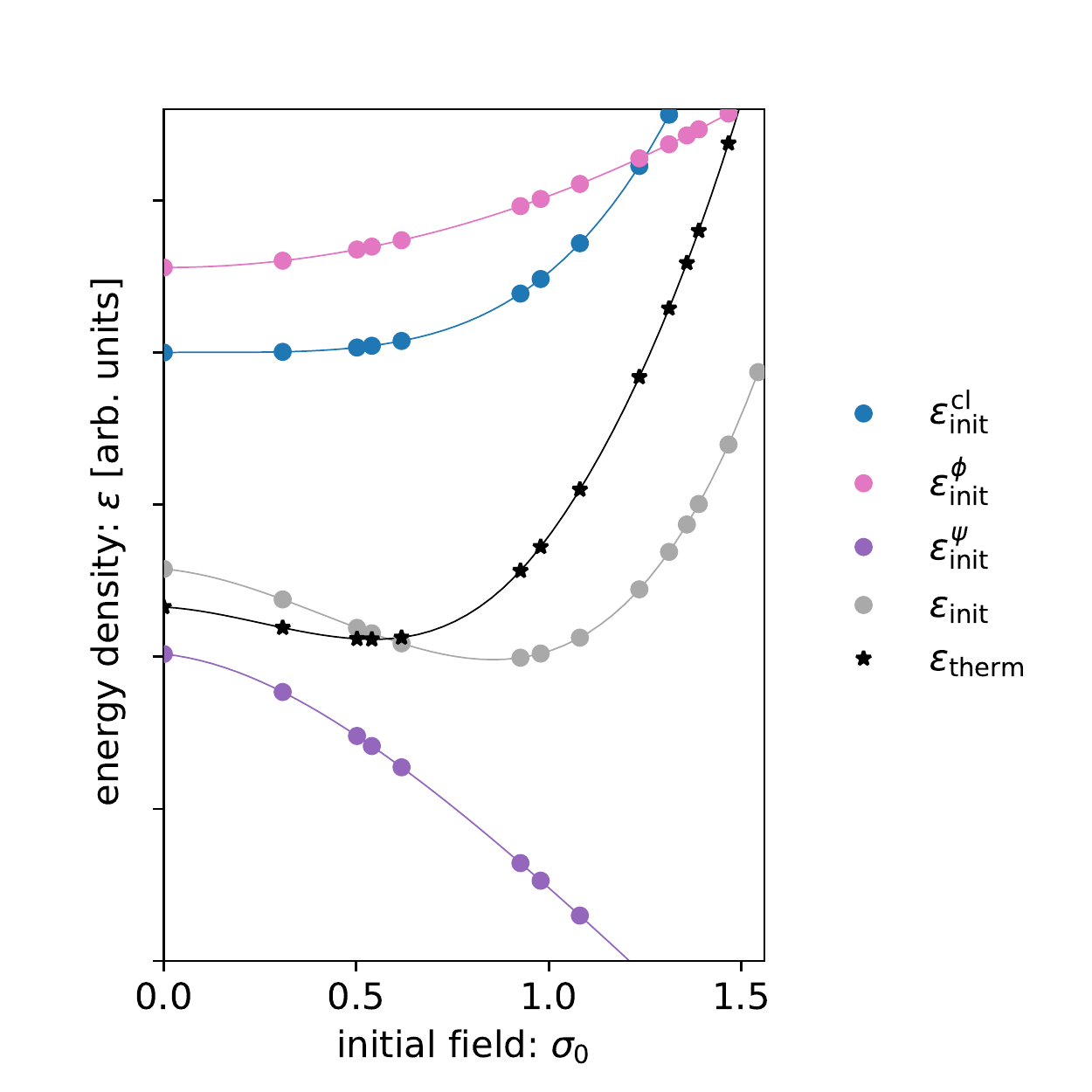}	
	\caption{
		The energy density at initial time $ \varepsilon_\mathrm{init} = \varepsilon (t=0) $ and at late times $ \varepsilon_\mathrm{th} = \overline{\varepsilon} $ as a function of the initial field value. 
		We present the classical, bosonic and fermionic contributions to the initial energy density separately. Together, they form a bounded shape with minimum at a nonzero initial field value (gray curve). 
		At late times, the energy density reaches the constant shape shown by $ \varepsilon_\mathrm{therm} $ in black. The minimum of the energy density at late times corresponds to the maximal field values found.
		The initial field is given in units of the pseudocritical temperature $ T_{pc} $ (cf.  Figure~\ref{fig:order_parameters}).
	}
	\label{fig:energy_density}
\end{figure*}

In order to study the influence of the initial field, we consider the energy density computed at initial time $ \varepsilon_\mathrm{init} $. In Figure~\ref{fig:energy_density} we show the contributions from the field, the bosons and the fermions separately. The blue and pink lines show, how the field and the bosonic energy densities exhibit a positive curvature. In contrast, the fermionic contribution shown in violet leads to a tilt toward negative curvature, which is a consequence of the explicit chiral symmetry breaking. Summing the three parts together, one obtains the total energy density that has a minimum at a nonzero field value, as represented by the gray curve. 

It is important to note that the energy density $ \varepsilon_\mathrm{init} $ computed at time $ t=0 $ does not include the quantum fluctuations that are generated dynamically by the field. As the field generates further fluctuations, the effective potential is pushed toward a more symmetric form with its minimum moving toward smaller field expectation values. In order to see this, we also look at the energy density computed at late times, where the system is thermalized. The result is shown by the black line in Figure~\ref{fig:energy_density}. It can be seen that the energy density indeed becomes steeper and the minimum moves toward smaller field values. Thus, the energy density provides a useful quantity in order to study the impact of quantum fluctuations on the effective potential, although we emphasize that the energy density and the effective potential are two different quantities. 

Having this qualitative picture of the effective quantum potential in mind, we can understand the behavior of the three curves shown in the left plot of Figure~\ref{fig:field_evolution_IC}. 
If the initial field sits close to the minimum of the effective potential, it barely oscillates and hence almost no additional fluctuations are created dynamically. Accordingly, the shape of the potential does not change with the time evolution such that the position of the minimum stays the same. This is shown by the black line. 
In contrast, the field can be placed on a point away from the potential minimum. As it starts moving toward the potential minimum, the field dynamically generates fluctuations. These fluctuations change the shape of the potential,  thereby altering the position of the minimum. The further away the field is from the potential minimum in the beginning, the more fluctuations are generated and the stronger the potential deforms. As we increase the distance of the initial field from the potential minimum at time $ t=0 $, the minimum of the potential at late times moves toward zero. Examples of this behavior are depicted by the green and red curves in the left plot Figure~\ref{fig:field_evolution_IC}.

As discussed in the previous section, energy cannot only be provided in terms of a nonzero initial field value (and the fluctuations this field generates), but also in terms of occupancies. 
Hence, the same late-time field value can be approached for different initial conditions. 
In the right plot of Figure~\ref{fig:field_evolution_IC} the time evolution of $ \braket{\sigma(t)} $ is shown for the two simulations discussed in Section~\ref{sec:intermediate_energy_densities}.  The blue line displays the time evolution of the field starting from field dominated initial conditions, while the orange line shows the time evolution starting from fluctuation dominated initial conditions. For both initial conditions the quantum potential has the same minimum, characterized by a late-time stationary field value of $\overline{\sigma} = 0.33 $. 

Although we commonly say that initial conditions with the same energy density lead to the same thermal state, there is a caveat. 
Two initial states thermalizing at the same late-time state usually do not have the same energy density at time $ t=0 $ because the energy density computed at initial time does not include dynamically generated fluctuations. What one means is that different initial conditions provide the same amount of fluctuations to the system. The way they are provided depends on the initial state and partly they are generated dynamically. 
However, for the quantum thermal equilibrium state that is approached at late times only the amount of fluctuations introduced to the system is relevant. 

\subsection{Thermal equilibrium}
\label{sec:equilibrium}
\subsubsection{Field expectation value}

After discussing the time evolution of the field expectation value, we now turn to its late-time properties. 
We denote the stationary value of the field at late times by $ \overline{\sigma} $. As discussed in \ref{sec:thermal_equilibrium}, at these times the fluctuation-dissipation theorem is satisfied and the system state is considered to be thermal. Thus, we consider $ \overline{\sigma} $ to be the thermal field expectation value. 

The late-time field values $ \overline{\sigma} $ are determined by the average of field values over a time range $ [t^*, t^* + \Delta t] $ with $ t^* $ being a time at which the field is sufficiently stationary. For the results shown in this work, we use $ t^* = 130 $ and $ \Delta t = 130 $, such that the standard deviation of the mean is $ \mathcal{O}(10^{-4}) $ to $ \mathcal{O}(10^{-11})$ depending on the initial conditions used.

\begin{figure*}[t]
	\centering
	\includegraphics[width=1.0\textwidth]{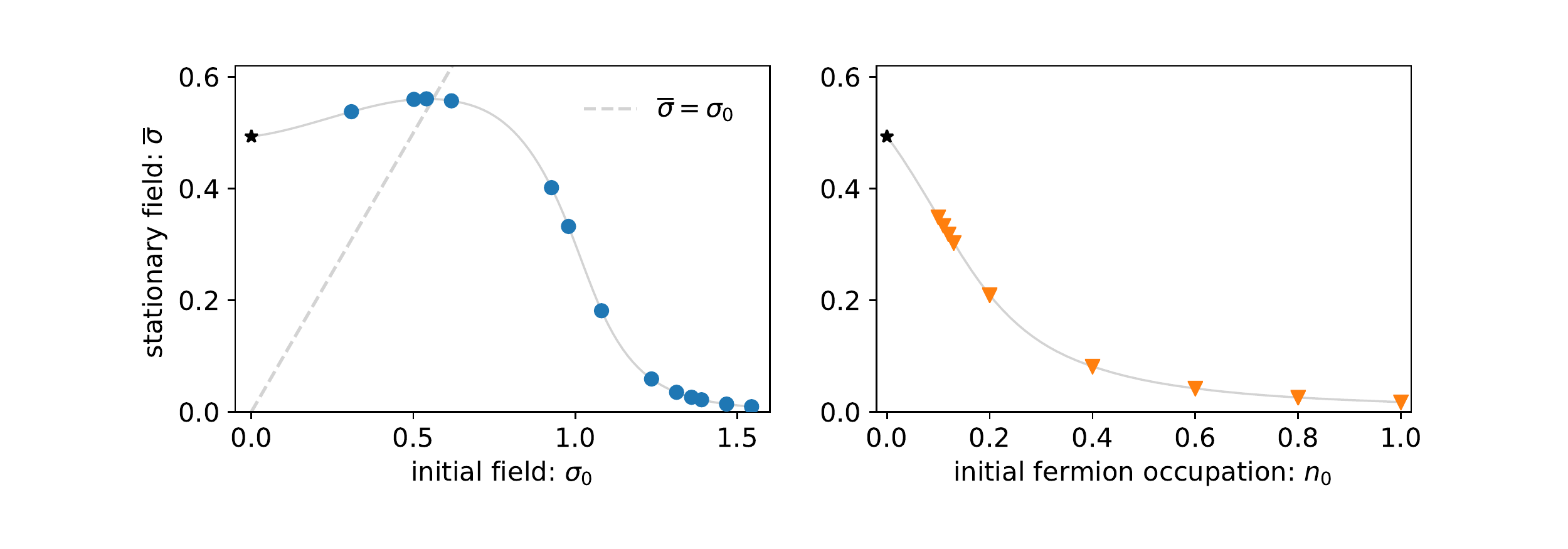}	
	\caption{
		The value of the thermalized one-point function for different initial conditions. On the right, the thermal field value $ \overline{\sigma} $ is shown for initial conditions with different field values $ \sigma_0 $. The gray dashed line indicates $ \overline{\sigma} = \sigma_0 $. 
		On the left, the thermal field value is shown for different initial fermion occupation numbers $ n_0 $. 
		In both plots, the black star indicates the value obtained for initial conditions with $ n_0 = 0 $ and $ \sigma_0 = 0 $. 
		In both plots gray lines show cubic spline interpolations of the data points.	Dimensionful quantities are given in units of the pseudocritical temperature $ T_{pc} $ (cf.  Figure~\ref{fig:order_parameters}).
	}
	\label{fig:field_value_IC}
\end{figure*}

First, let us look at the time evolution of the macroscopic field for different initial field values $ \sigma_0 $. Naively, one might expect that larger field values automatically imply increasing energy densities in the initial state, hence a higher thermalization temperature and smaller field value. However, as the discussion above already pointed out, this is not the case. In the left plot of Figure~\ref{fig:field_value_IC} we show how the late-time field value $ \overline{\sigma}$ depends on the initial field $ \sigma_0 $. With increasing $ \sigma_0 $, the thermal field $ \overline{\sigma} $ first grows and then decays to zero. 
The maximal value for $ \overline{\sigma} $ is expected, when the least amount of fluctuations is generated dynamically, as these fluctuations would push the minimum of the potential and thus $ \overline{\sigma} $ toward zero. We indeed find the largest late-time field values for $ \sigma_0 \approx \overline{\sigma}$, which in indicated by the gray dashed line in the plot. 

Second, we consider fluctuation dominated initial conditions where the field value is set to $ \sigma_0 = 0 $ while the initial fermion occupation is taken to be constant, i.e. $ n_\psi(t=0, |\mathbf{p}|) = n_0 $, and varied between zero and one. 
In the right plot of Figure~\ref{fig:field_value_IC} we can see that increasing the fermion occupation number $ n_0 $ goes along with smaller thermal field values $ \overline{\sigma}  $. Thus, for larger $ n_0 $ higher temperatures are reached, emphasizing again that larger fermion occupation numbers lead to a rise of the fluctuations that make the effective potential more symmetric. 

\begin{figure*}[t]
	\centering
	\includegraphics[width=.5\textwidth]{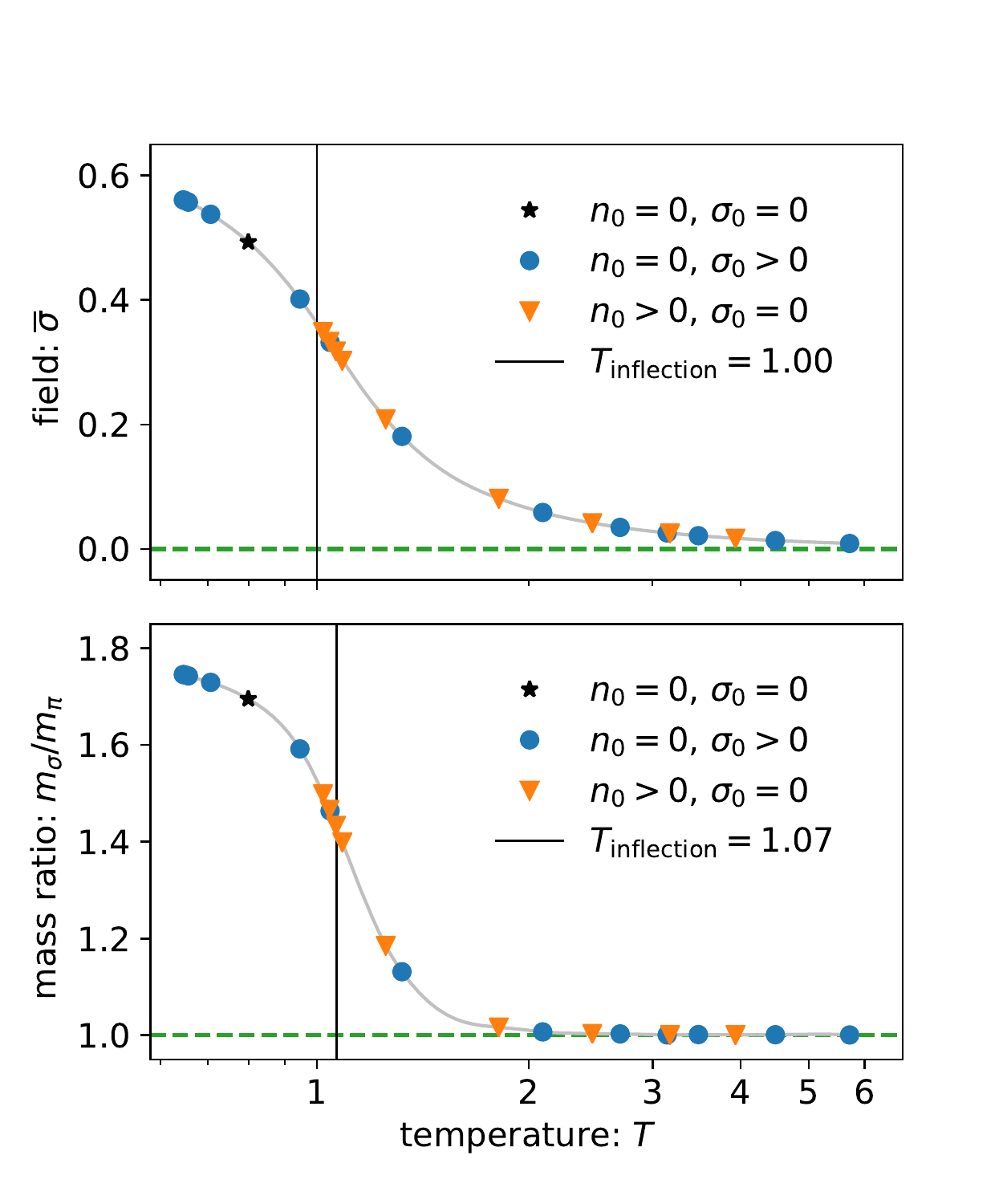}	
	\caption{
		Order parameters of the quark-meson model as a function of temperature. In the upper plot, the order parameter is given by the macroscopic field $ \overline{\sigma} $ which is the thermalized value of the one-point function. 
		In the lower plot, the order parameter is given by the ratio of the $ \sigma $ meson and pion masses. The masses are derived from the two-point functions of the corresponding bosonic fields. 
		The gray lines show cubic spline fits to the data points. The inflection points are indicated by the black vertical lines. 
		Dimensionful quantities are given in units of the pseudocritical temperature $ T_{pc} $ defined as the inflection point $ T_\mathrm{inflection} $ of the order parameter $ \overline{\sigma} $ shown in the upper plot. 
		}
	\label{fig:order_parameters}
\end{figure*}

\subsubsection{Crossover phase transition}

Our results regarding the thermal state of the system can be summarized in an analysis of the crossover transition between the chiral broken and the chiral symmetric phase of the quark-meson model. When a system becomes thermal, the thermodynamic concept of a phase diagram can be applied. The conjectured phase diagram of the quark-meson model contains important features of the QCD phase diagram. It exhibits a chiral symmetric phase with vanishing field expectation value at high temperature $ T $, as well as a chiral broken phase with nonzero field corresponds at low $ T $ .

In order to study the phase transition and the transition temperature, we employ two different order parameters, one deduced from the one-point function and one from the two-point functions. 
The first one is the field expectation value of the thermalized field $ \overline{\sigma} $. It is nonzero in the chirally broken phase and zero in the chirally symmetric phase. Often this field value is identified as the pion decay constant $ f_\pi $. 
The second one is the mass ratio $ m_\sigma / m_\pi$, where $m_ \sigma $ and  $m_\pi $ are the masses of the $ \sigma $ meson and the pion, respectively. The masses are determined from the bosonic spectral functions as discussed in Section~\ref{sec:equilibrium_spectra_mesons}. In the chiral limit, the mass ratio is expected to go to unity. 

Starting from the different initial states analyzed, we find that the system thermalizes at different temperatures. Thereby, the dependence of an order parameter on the temperature provides insight into the nature of the phase transition. 
In Figure~\ref{fig:order_parameters} we show our numerical results for the temperature dependence of the two order parameters, $ \overline{\sigma} $ in the upper and $ m_\sigma / m_\pi$ in the lower plot. 
Every point in the diagram corresponds to a simulation with a different initial state. As indicated in the legend, we are considering initial states of various fermion occupations, described by $ n_0 $, and initial field values, described by $ \sigma_0 $. It is reassuring to see that the order parameters obtained from field or fluctuation dominated initial conditions align themselves on a single curve, which is characteristic for a smooth crossover transition. This is yet another way of seeing that the thermal states are independent of the details of the initial conditions. 

As chiral symmetry is restored with rising temperature, the field value decays to zero while the mass ratio goes down to one. 
The field expectation value $ \overline{\sigma} $ is often considered as a first approximation for the pion decay constant $ f_\pi $. As can be seen, in the limit $ T\rightarrow 0 $ some value $ \overline{\sigma}  \simeq \mathcal{O}(1)$ is approached. At the lowest temperature considered we find $ \overline{\sigma} / m_\pi \simeq 0.65 $, matching the phenomenological value $ f_\pi / m_\pi \simeq 0.69 $ \cite{Tanabashi:2018oca}. 
Further, we can see from the lower plot in Figure~\ref{fig:order_parameters} that the mass ratio is only $ m_\sigma/m_\pi \simeq 1.8 $ at the lowest temperatures available, which is smaller than the expectation from the known values of the masses. However, the mass ratio is expected to further increase with decreasing temperature. 

We perform a cubic spline fit to the data points and identify the inflection point of the field $ \overline{\sigma} $ as the pseudocritical temperature of the crossover $ T_{pc} $. We indicate the inflection point of both the field and the mass ratio in the plots of Figure~\ref{fig:order_parameters}. It can be seen that the temperatures deduced from the two different order parameters are comparable with each other. We find that the pseudocritical temperature is of the order of the pion mass. This is in agreement with the expectation of the QCD phase transition being at around $  \SI{150}{MeV}$ for vanishing baryon density. 

\subsection{Spontaneous symmetry breaking}
\label{sec:SSB}

\begin{figure*}[t]
	\centering
	\includegraphics[width=1.\textwidth]{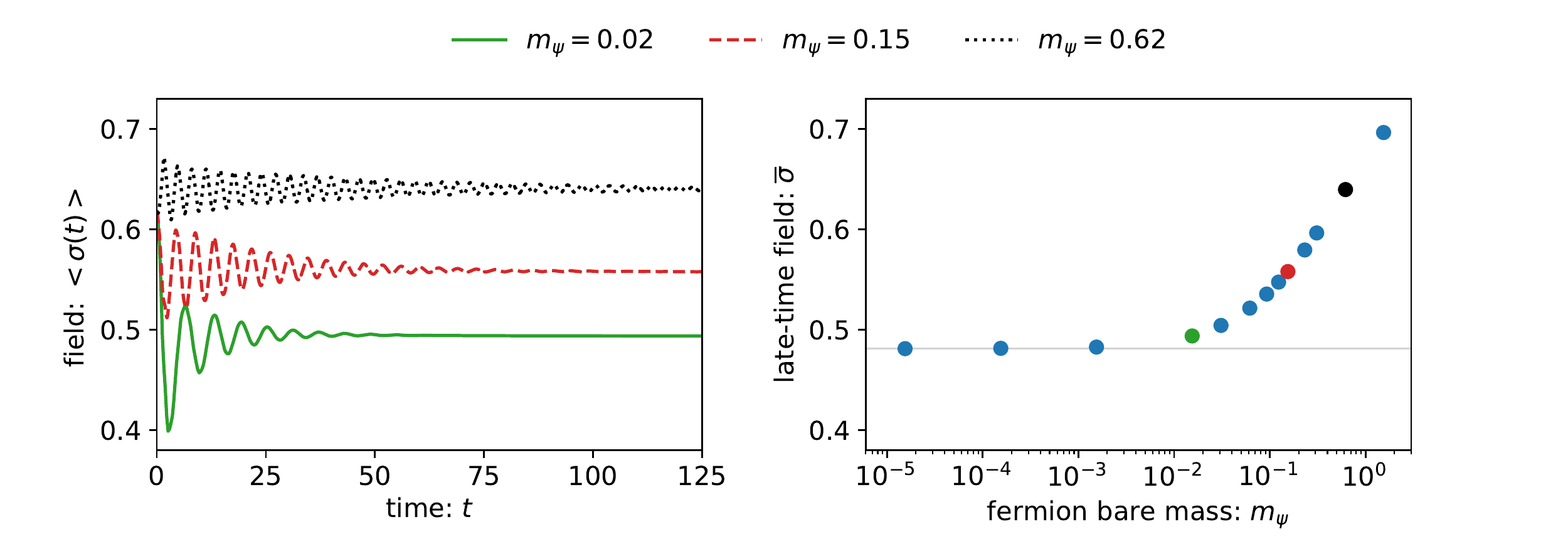}	
	\caption{
		Left: The time evolution of the field with initial value $ \sigma_0 = 0.62 $ shown for three different bare fermion masses $ m_\psi $. The field reaches the stationary value $ \overline{\sigma} $ at late times. 
		Right: The asymptotic field value $ \overline{\sigma} $ shown for different bare fermion masses $ m_\psi $. The green, red and black data points correspond to the simulations shown in the left plot. The field value decreases with the fermion bare mass and approaches an asymptotic value for $ m_\psi \rightarrow 0 $.
		Dimensionful quantities are given in units of the pseudocritical temperature $ T_{pc} $ (cf.  Figure~\ref{fig:order_parameters}).
	}
	\label{fig:field_SSB}
\end{figure*}

We have seen that the explicit chiral symmetry breaking in the system leads to nonzero field expectation values. Here, we analyze the limit of vanishing explicit symmetry breaking, i.e. $ m_\psi \rightarrow 0 $, with spontaneous symmetry breaking still present. 

If the fermion bare mass vanishes, i.e., $ m_\psi = 0 $, the action of the quark-meson model \eqref{eq:action} is invariant under chiral $ SU_L(2) \times SU_R (2) \sim O(4)$ transformations and therefore symmetric under chiral symmetry. 
Still, a nonzero field expectation value can break this symmetry spontaneously. 
For nonzero fermion bare masses, chiral symmetry is explicitly broken and the minimum of the potential is located at some nonzero field value. If the field expectation value stays nonzero for $ m_\psi \rightarrow 0$, we expect to observe spontaneous symmetry breaking. 

We compare simulations with different fermion bare masses $ m_\psi  $ while all other parameters of the theory are kept fixed. 
The system is studied for initial conditions with $ \sigma_0 =  0.62 $ and vanishing fermion and boson occupations, i.e. $ n_0 = 0 $. 
In the left plot of Figure~\ref{fig:field_SSB}, we show the time evolution of the field expectation value $ \braket{\sigma(t)} $, for three examples with different fermion bare masses. As before, the field oscillates before it equilibrates to the thermal late-time value $ \overline{\sigma} $. 

Since the fermion bare mass $ m_\psi $ governs the strength of the chiral symmetry breaking and thus the deformation of the potential, increasing fermion bare masses yields larger values for $ \overline{\sigma} $. At the same time $ m_\psi $ determines the fermion backreaction on the field, i.e. how strong the field is pushed away from its current value. The field only reaches a stationary value, if the backraction from the fermions on the field and the bosonic interactions with the field balance out. 

Here, fermion bare masses with values from $ \mathcal{O}(10^{-4}) $ ranging to $ \mathcal{O}(1) $ are considered. 
We find that the field approaches the asymptotic value $ \overline{\sigma} = 0.48 $ for $ m_\psi \rightarrow 0 $, which is shown in the right plot of Figure~\ref{fig:field_SSB}. 
This analysis shows that our numerical simulations of the quark-meson model reproduce the expected spontaneous symmetry breaking in the limit of vanishing fermion bare mass.

\section{Summary and conclusion}
\label{sec:conclusion}

Motivated by current experimental studies of the QCD phase diagram in heavy-ion collisions, we investigated the dynamical approach of the quark-meson model to thermal equilibrium using a range of different initial conditions dominated by either the sigma field or fermionic fluctuations. The time evolution of one- and two-point functions was computed numerically using closed equations of motion derived from the 2PI effective action at NLO in $1/N$ and the Yukawa coupling. 

We show that our simulations correctly capture the approach to thermal equilibrium, which depends only on the energy density of the initial condition. The crossover phase transition from the chiral broken phase at low temperatures to the chiral symmetric phase at high temperatures is reproduced by the late-time equilibrium states. Thermalization in the chiral broken phase is characterized by a finite field expectation value, a mass difference between the sigma meson and the pions as well as narrow quasiparticle peaks in the spectrum. The restoration of chiral symmetry in the high-temperature regime expresses itself in the field expectation value decreasing to zero, the mass ratio of $ \sigma $ and $ \pi $ mesons going to unity and the scalar component of the quark spectral functions decaying to zero. 

Our investigation focused in detail on the dynamical thermalization revealing differences in the time evolution depending on the initial state employed.
We not only studied the time evolution of the field expectation value but also probed the dynamical properties of the two-point functions, expressed in terms of the spectral and statistical functions, which carry information about the available quasiparticle states and their occupation in the system, respectively. 
For initial states with vanishing initial field but energy supplied by fermion occupation, the spectral and statistical functions of both quarks and mesons approach their late-time thermal shapes already at early times. 
In contrast, if the energy density is predominantly provided by the nonzero initial field value, the redistribution of energy from the field first to the bosonic sector and subsequently to the fermionic sector leads to high occupancies of the mesons at intermediate stages. 
This is also reflected by the different behavior found in the time evolution of the  quasiparticle masses depending on the initial conditions. 

The deployed nonequilibrium setup of the quark-meson model captures important features of the low-energy behavior of QCD. 
By studying the temperature dependence of the quasiparticle masses, we find that the lightest degrees of freedom are given by the pions at temperatures below and by quarks above the phase transition. This implies that quarks are the relevant degrees of freedom at high temperatures while pions dominate below the critical temperature. 
Furthermore, we learn from the width that at high temperatures the more energetic high-momentum decay modes are more pronounced than for low temperatures. 

The nonvanishing expectation value of the sigma field describes the order parameter of the chiral phase transition. Its dynamics depends on the initial state. If the initial field value is close to the minimum of the effective potential, the field remains almost constant. Otherwise, the field rolls down a potential hill and starts oscillating, thereby dynamically generating fluctuations. 

Having shown that the dynamics of the thermalization process reveal interesting features before approaching the final thermal state, we lay the foundation for future investigations of the quark-meson model with nonzero baryon chemical potential. 
In particular, the possibility of probing the dynamical thermalization of systems surpassing the critical point of the chiral phase transition is of outermost interest.

\section*{Acknowledgments}

The authors acknowledge support by the state of Baden-W\"urttemberg through bwHPC and the German Research Foundation (DFG) through the Collaborative Research Centre ``SFB 1225 (ISOQUANT)". 

A. R. acknowledges funding from the Research Council of Norway under the FRIPRO Young Research Talent Grant No. 286883 and Grant No. 295310. This work has utilized computing resources provided by UNINETT Sigma2 - the National Infrastructure for High Performance Computing and Data Storage in Norway under project NN9578K-QCDrtX ``Real-time dynamics of nuclear matter under extreme conditions". 

\begin{widetext}
\appendix
\section{2PI equations}
\label{app:eom}

In this appendix we review the equations of motion for the macroscopic field as well as for the bosonic and fermionic propagators obtained from the 2PI effective action given in  \eqref{eq:2PIeffectiveAction_QMM}. For a derivation of the exact evolution equations of the bosonic and fermionic equations of motion we refer to \cite{Aarts:2002dj,Berges:2001fi} and \cite{Berges:2002wr}, respectively. 

Quantum corrections are included in the evolution equations via the self-energy terms. 
In the bosonic sector, we take into account all leading and subleading quantum corrections in an expansion of the 2PI effective action in $ 1/N $, where $ N=4 $ is the number of bosonic field components. The corresponding 2PI diagrams are depicted in the first two rows of Figure~\ref{fig:Diagrams}. The self-energy expressions for the bosonic quantum corrections have been derived in \cite{Berges:2001fi}. 
Additionally, we include the NLO in the Yukawa coupling $ g $ corresponding to the contribution coming from a fermion-boson loop; see the third row in Figure~\ref{fig:Diagrams}. 
The self-energy terms for the chirally symmetric case of $ m_\psi =0$ have been derived in \cite{Berges:2002wr,Berges:2009bx}. For nonvanishing fermion bare mass $ m_\psi $ a computation of the self-energy contributions can be found in \cite{tuprints2209}. For completeness, we provide a summary of all relevant equations for the time evolution.

\subsection{Exact evolution equations}

We first outline the exact evolution equations---for the macroscopic field, the bosonic as well as the fermionic two-point functions---without any approximations but including the assumption of symmetries. These equations are derived from first principles and capture the whole quantum evolution. As they are too complicated to be solved analytically without approximations, we employ a truncation of the 2PI effective action which yields the self-energies presented subsequently. 

We note that the bosonic and fermionic evolution equations are causal, meaning that they only depend on times prior to the evaluation. The history of the time evolution is contained in \textit{memory integrals} \cite{Berges2004}.

The equation of motion for the macroscopic field is determined by the stationary conditions of the effective action. 
Since the expectation values of the pion fields are zero, the only relevant equation is the one for $ \phi_1(t) \equiv \sigma(t) $. Making use of the decomposition into spectral and statistical functions, the field equation in absence of external source terms can be written as
\begin{align}
\left[	
\partial_t^2 + m^2 + \dfrac{\lambda}{6N} \sigma^2(t) + \dfrac{\lambda}{6N} \left( 3 \int _\mathbf{p} F_\sigma(t, t, \mathbf{p}) + (N-1) \int _\mathbf{p} F_\pi (t, t, \mathbf{p}) \right)
\right]  \sigma(t)
=	4g \int _\mathbf{p} F_S(t, t, \mathbf{p}) 
+ \dfrac{\delta \Gamma_2[G, \Delta, \phi] }{\delta\sigma(t)}\,.
\label{eq:eom_field}
\end{align}
In this expression the tadpole terms with $ F_\sigma $ and $ F_\pi $ originate from one-loop corrections and are therefore independent of any truncation in the 2PI effective action. The $ F_S $ term corresponds to a source term representing the backreaction of the fermions on the field. The functional $ \Gamma_2[G, \Delta, \phi] $ contains all contributions from 2PI vacuum diagrams. Hence, the $ \Gamma_2 $ term describes all nonlocal contributions to the interactions of the spectral and statistical functions with the field $ \phi $. 

For both quarks and mesons it is advantageous to rewrite the relevant equations of motion in terms of the statistical and spectral components. The time-ordered two-point functions $ G $ and $ \Delta $ are related to the statistical and spectral functions according to 
\begin{align}\nonumber 
G_{ab}(x,y) &= F^\phi_{ab} (x,y)- \dfrac{i}{2} \rho^\phi_{ab}(x,y)  \sgn _\mathcal{C} (x^0 - y^0), \\
\Delta_{AB}(x,y) &= F^\psi_{AB} (x,y)- \dfrac{i}{2} \rho^\psi_{AB}(x,y)  \sgn _\mathcal{C}(x^0 - y^0), 
\end{align}
where the sign function is taken along the closed time path. Equivalently, the self-energies can be decomposed into statistical and spectral components, 
\begin{align}
\Sigma_i(x,y) 
&= 
C_i (x,y)
- \dfrac{i}{2} A_i(x,y)  \sgn _\mathcal{C} (x^0 - y^0) 	- i \Sigma^{\text{local}}_i (x) \delta(x-y) \,, 
\end{align}
where $ i = \sigma, \pi $ in the bosonic and $ i = S, 0, V, T $ in the fermionic case. The local contribution $ \Sigma^{\text{local}} $ only appears in the bosonic sector for the considered model. In this notation, $ C $ and $ A $ describe the symmetric and antisymmetric components of the self-energy terms. 
Since the numerical calculations are performed in spatial momentum space, we choose to present the formulas in terms of Fourier modes and refer to \cite{tuprints2209} for details on the calculation. The complete set of evolution equations for the bosonic sector is given by 
	\begin{align} \nonumber
	\left[
	\partial_t^2 + \mathbf{p}^2 + M^2_i(t)
	\right]
	F_i(t, t', |\mathbf{p}|)
	&=
	- \int_{t_0}^t \diff t''	A_i(t, t'', |\mathbf{p}| ) F_i(t'', t', |\mathbf{p}|)\\\nonumber  & \hspace{.4cm}
	+ \int_{t_0}^{t'} \diff t''	C_i(t, t'', |\mathbf{p}| ) \rho_i(t'', t', |\mathbf{p}|), \\[2ex]
	\left[
	\partial_t^2 + \mathbf{p}^2 + M^2_i(t)
	\right]
	\rho_i(t, t', |\mathbf{p}|)
	&=
	- \int_t^{t'} \diff t''	A_i(t, t'', |\mathbf{p}| ) \rho_i(t'', t', |\mathbf{p}|) \,,
	\end{align}
where $ i = \sigma, \pi $ and the time-dependent masses read 
\begin{align}\nonumber 
M_\sigma^2(t) &= 
m^2 + \dfrac{\lambda}{2 N} \sigma^2(t)  + \Sigma_\sigma^{\text{local}} (t)\,,\\[1ex]
M_\pi^2(t) &= 	
m^2 + \dfrac{\lambda}{6N} \sigma^2(t)  + \Sigma_\pi^{\text{local}} (t)\,,
\end{align}
which are also referred to as \textit{gap equations}. 
The coupling between $ \sigma $ and $ \pi $ components of the bosonic two-point functions solely occurs via the self-energies. 

For the fermions we take into account the four components introduced in the main text: scalar, vector-zero, vector and tensor. The evolution equations for the fermionic statistical propagators are

\begingroup
\allowdisplaybreaks
\begin{subequations}
	\begin{align}
	i \partial_t F_S(t,t', |\mathbf{p}|)
	&=  
	-i |\mathbf{p}| \ F_T (t,t', |\mathbf{p}|)
	+ M_\psi(t)  F_0(t,t', |\mathbf{p}|) \nonumber\\
	&\quad
	+ \int_{t_0} ^t \diff t ''
	\Big[
	A_S(t,t'', |\mathbf{p}|) \ F_0(t'',t', |\mathbf{p}|)
	+ A_0(t,t'', |\mathbf{p}|) \ F_S (t'',t', |\mathbf{p}|)\nonumber\\
	& \quad \qquad \qquad \quad
	+ i A_V(t,t'', |\mathbf{p}|) \ F_T (t'',t', |\mathbf{p}|) 
	- i A_T(t,t'', |\mathbf{p}|)  \ F_V(t'', t', |\mathbf{p}|)
	\Big] \nonumber\\
	&\quad
	+ \int_{t_0}^{t'} \diff t ''
	\Big[
	- C_S(t,t'', |\mathbf{p}|) \ \rho_0(t'',t', |\mathbf{p}|)
	- C_0(t,t'', |\mathbf{p}|) \ \rho_S (t'',t', |\mathbf{p}|)\nonumber\\
	& \quad \qquad \qquad \quad
	- i C_V(t,t'', |\mathbf{p}|) \ \rho_T (t'',t', |\mathbf{p}|) 
	+ i C_T(t,t'', |\mathbf{p}|)  \ \rho_V(t'', t', |\mathbf{p}|)
	\Big], \\[2em]
	i\partial_t F_0(t,t', |\mathbf{p}|)
	&=
	|\mathbf{p}| F_V(t,t', |\mathbf{p}|)
	+ M_\psi(t) F_S(t,t', |\mathbf{p}|)\nonumber\\
	&\quad+ \int_{t_0} ^t \diff t ''
	\Big[
	A_S(t,t'', |\mathbf{p}|) \ F_S(t'',t', |\mathbf{p}|)
	- A_T(t,t'', |\mathbf{p}|) \ F_T (t'',t', |\mathbf{p}|)\nonumber\\
	& \quad \qquad \qquad \quad
	+  A_0(t,t'', |\mathbf{p}|) \ F_0 (t'',t', |\mathbf{p}|) 
	- A_V(t,t'', |\mathbf{p}|)  \ F_V(t'', t', |\mathbf{p}|)
	\Big] \nonumber\\
	&\quad+ \int_{t_0} ^{t'} \diff t ''
	\Big[
	-C_S(t,t'', |\mathbf{p}|) \ \rho_S(t'',t', |\mathbf{p}|)
	+ C_T(t,t'', |\mathbf{p}|) \ \rho_T (t'',t', |\mathbf{p}|)\nonumber\\
	& \quad \qquad \qquad \quad
	- C_0(t,t'', |\mathbf{p}|) \ \rho_0 (t'',t', |\mathbf{p}|) 
	+ C_V(t,t'', |\mathbf{p}|)  \ \rho_V(t'', t', |\mathbf{p}|)
	\Big], \\[2em]		
	\partial_t F_V(t,t', |\mathbf{p}|)
	&=
	-i |\mathbf{p}| F_0(t,t', |\mathbf{p}|)
	+ M_\psi(t) F_T(t,t', |\mathbf{p}|)\nonumber\\
	&\quad+ \int_{t_0} ^t \diff t ''
	\Big[
	-i A_0(t,t'', |\mathbf{p}|) \ F_V(t'',t', |\mathbf{p}|)
	+ i A_V(t,t'', |\mathbf{p}|) \ F_0 (t'',t', |\mathbf{p}|)\nonumber\\
	& \quad \qquad \qquad \quad
	+  A_S(t,t'', |\mathbf{p}|) \ F_T (t'',t', |\mathbf{p}|) 
	+ A_T(t,t'', |\mathbf{p}|)  \ F_S(t'', t', |\mathbf{p}|)
	\Big] \nonumber\\	
	&\quad+ \int_{t_0}^{t'}  \diff t ''
	\Big[
	+ i C_0(t,t'', |\mathbf{p}|) \ \rho_V(t'',t', |\mathbf{p}|)
	- i C_V(t,t'', |\mathbf{p}|) \ \rho_0 (t'',t', |\mathbf{p}|)\nonumber\\
	& \quad \qquad \qquad \quad
	-  C_S(t,t'', |\mathbf{p}|) \ \rho_T (t'',t', |\mathbf{p}|) 
	- C_T(t,t'', |\mathbf{p}|)  \ \rho_S(t'', t', |\mathbf{p}|)
	\Big], \\[2em]	
	\partial_t F_T(t,t', |\mathbf{p}|)
	&=
	|\mathbf{p}| F_S(t,t', |\mathbf{p}|)
	- M_\psi(t) F_V(t,t', |\mathbf{p}|)\nonumber\\
	&\quad+ \int_{t_0} ^t \diff t ''
	\Big[
	- A_S(t,t'', |\mathbf{p}|) \ F_V(t'',t', |\mathbf{p}|)
	- A_V(t,t'', |\mathbf{p}|) \ F_S (t'',t', |\mathbf{p}|)\nonumber\\
	& \quad \qquad \qquad \quad
	-i A_0(t,t'', |\mathbf{p}|) \ F_T (t'',t', |\mathbf{p}|) 
	+ iA_T(t,t'', |\mathbf{p}|)  \ F _0(t'', t', |\mathbf{p}|)
	\Big] \nonumber\\
	&\quad+ \int_{t_0}^{t'}  \diff t ''
	\Big[
	+ C_S(t,t'', |\mathbf{p}|)  \ \rho_V(t'',t', |\mathbf{p}|)
	+ C_V(t,t'', |\mathbf{p}|)  \ \rho_S (t'',t', |\mathbf{p}|)\nonumber\\
	& \quad \qquad \qquad \quad
	+ i C_0(t,t'', |\mathbf{p}|)  \ \rho_T (t'',t', |\mathbf{p}|) 
	- iC_T(t,t'', |\mathbf{p}|)  \  \rho _0(t'', t', |\mathbf{p}|)
	\Big],
	\end{align}
	\label{eq:eom_fermionF}%
\end{subequations}
\endgroup
and for the fermionic spectral functions 
\begin{subequations}
	\begin{align}
	i \partial_t \rho_S(t,t', |\mathbf{p}|)
	&=  
	-i |\mathbf{p}| \ \rho_T (t,t', |\mathbf{p}|)
	+ M_\psi(t)  \rho_0(t,t', |\mathbf{p}|) \nonumber\\
	&\quad
	+ \int_{t'} ^t \diff t ''
	\Big[
	A_S(t,t'', |\mathbf{p}|) \ \rho_0(t'',t', |\mathbf{p}|)
	+ A_0(t,t'', |\mathbf{p}|) \ \rho_S (t'',t', |\mathbf{p}|)\nonumber\\
	& \quad \qquad \qquad \quad
	+ i A_V(t,t'', |\mathbf{p}|) \ \rho_T (t'',t', |\mathbf{p}|) 
	- i A_T(t,t'', |\mathbf{p}|)  \ \rho_V(t'', t', |\mathbf{p}|)
	\Big], \\[2em]
	i\partial_t \rho_0(t,t', |\mathbf{p}|)
	&=
	|\mathbf{p}| \rho_V(t,t', |\mathbf{p}|)
	+ M_\psi(t) \rho_S(t,t', |\mathbf{p}|)\nonumber\\
	&\quad+ \int_{t'} ^t \diff t ''
	\Big[
	A_S(t,t'', |\mathbf{p}|) \ \rho_S(t'',t', |\mathbf{p}|)
	- A_T(t,t'', |\mathbf{p}|) \ \rho_T (t'',t', |\mathbf{p}|)\nonumber\\
	& \quad \qquad \qquad \quad
	+  A_0(t,t'', |\mathbf{p}|) \ \rho_0 (t'',t', |\mathbf{p}|) 
	- A_V(t,t'', |\mathbf{p}|)  \ \rho_V(t'', t', |\mathbf{p}|)
	\Big], \\[2em]
	\partial_t \rho_V(t,t', |\mathbf{p}|)
	&=
	-i |\mathbf{p}| \rho_0(t,t', |\mathbf{p}|)
	+ M_\psi(t) \rho_T(t,t', |\mathbf{p}|)\nonumber\\
	&\quad+ \int_{t'} ^t \diff t ''
	\Big[
	-i A_0(t,t'', |\mathbf{p}|) \ \rho_V(t'',t', |\mathbf{p}|)
	+ i A_V(t,t'', |\mathbf{p}|) \ \rho_0 (t'',t', |\mathbf{p}|)\nonumber\\
	& \quad \qquad \qquad \quad
	+  A_S(t,t'', |\mathbf{p}|) \ \rho_T (t'',t', |\mathbf{p}|) 
	+ A_T(t,t'', |\mathbf{p}|)  \ \rho_S(t'', t', |\mathbf{p}|)
	\Big], \\[2em]
	\partial_t \rho_T(t,t', |\mathbf{p}|)
	&=
	|\mathbf{p}| \rho_S(t,t', |\mathbf{p}|)
	- M_\psi(t) \rho_V(t,t', |\mathbf{p}|)\nonumber\\
	&\quad+ \int_{t'} ^t \diff t ''
	\Big[
	- A_S(t,t'', |\mathbf{p}|) \ \rho_V(t'',t', |\mathbf{p}|)
	- A_V(t,t'', |\mathbf{p}|) \ \rho_S (t'',t', |\mathbf{p}|)\nonumber\\
	& \quad \qquad \qquad \quad
	-i A_0(t,t'', |\mathbf{p}|) \ \rho_T (t'',t', |\mathbf{p}|) 
	+ iA_T(t,t'', |\mathbf{p}|)  \ \rho _0(t'', t', |\mathbf{p}|)
	\Big]\,,
	\end{align}
	\label{eq:eom_fermion_rho}%
\end{subequations}
where the effective fermion mass is given by 
\begin{align}
M_\psi(x)
&= m_\psi + h \sigma(x). 
\end{align}
We note that the factors of $ i $ are necessary because the vector-zero component is imaginary while all other components are real. 

\subsection{Approximation Scheme}

At NLO in the large $ N $ expansion, the nonlocal interaction terms in the field equation \eqref{eq:eom_field} are given by 
\begin{align}
\dfrac{\delta\Gamma_2[\phi, G,\Delta]}{\delta\sigma(t)}
&=	
\dfrac{\lambda}{3N} \int_{t_0}^t \diff t' \int_\mathbf{p}
\ \Big[
I_\rho(t, t', |\mathbf{p}|)
F_\sigma(t, t', |\mathbf{p}|)
+ I_F (t, t', |\mathbf{p}|)
\rho_\sigma (t, t', |\mathbf{p}|)
\Big]\  \sigma(t')\,,
\label{eq:eom_phi_NLO}
\end{align}
where $ I_F $ and $ I_\rho $ are the spectral and statistical components of the summation functions presented below in \eqref{eq:summation_functions}. 

In our approximation, the local parts of the bosonic self-energies are
	\begin{align}\nonumber 
	\Sigma_\sigma^{\text{local}} (t; F) 
	&= 
	\dfrac{\lambda}{6N} 
	\int_\mathbf{q}
	\bigg[
	3 F_{\sigma }(t, t, |\mathbf{q}|)  + (N-1) F_\pi(t, t, |\mathbf{q}|)
	\bigg],\\[1ex]
	\Sigma_\pi^{\text{local}} (t; F) 
	&= 
	\dfrac{\lambda}{6N} 
	\int_\mathbf{q}
	\bigg[\ 
	F_\sigma(t, t, |\mathbf{q}|)  + (N+1) F_\pi(t, t, |\mathbf{q}|)
	\bigg].
	\end{align}
Since the nonlocal self-energy contributions form convolutions in momentum space, it is easier to evaluate them in coordinate space where they can be calculated as direct products. The relevant expressions for the boson self-energies in coordinate space are
\begin{subequations}
	\begin{align}
	C_{ab}^\phi(x,y) 
	&=
	- \dfrac{\lambda}{3N} 
	\Bigg\{
	I_F(x,y)
	\Big[
	\phi_a(t)\phi_b(t') + F_{ab}(x,y)
	\Big] 
	- \dfrac{1}{4} I_\rho(x,y)  \rho_{ab} (x,y) \nonumber
	+ P_F(x,y)
	F_{ab}(x,y)  
	- \dfrac{1}{4} P_\rho(x,y)
	\rho_{ab}(x,y)
	\Bigg\}
	\\\nonumber
	& \quad
	- 4 h^2 N_f \delta_{ab} 
	\Bigg\{
	F^\mu_V(x,y) F_{V, \mu} (x,y)
	- F_S (x,y) F_S(x,y)
	- F_T^{0i} (x,y) F_T^{0i}(x,y)	\\
	& \qquad \qquad \qquad \qquad \ 
	- \frac{1}{4}
	\bigg[
	\rho^\mu_V(x,y) \rho_{V, \mu} (x,y)
	- \rho_S (x,y)\rho_S(x,y)
	- \rho_T^{0i} (x,y) \rho_T^{0i}(x,y)
	\bigg]
	\Bigg\} , \\
	A_{ab}^\phi(x,y) 
	&=
	- \dfrac{\lambda}{3N} 
	\Bigg\{
	I_\rho(x,y)
	\Big[
	\phi_a(t)\phi_b(t') + F_{ab}(x,y)
	\Big] 
	+ I_F(x,y)  \rho_{ab} (x,y)
	+ P_\rho(x,y) F_{ab}(x,y)  
	+ P_F(x,y)
	\rho_{ab}(x,y)
	\Bigg\}\nonumber\\
	& \quad
	- 8 h^2 N_f \delta_{ab} 
	\Bigg\{
	\rho^\mu_V(x,y) F_{V, \mu} (x,y)
	- \rho_S(x,y) F_S(x,y)
	- \rho_T^{0i} (x,y)F_T^{0i}(x,y)
	\Bigg\}, 
	\end{align} \end{subequations}
where spatial isotropy and homogeneity are implied. 
The statistical and spectral components of the summation functions are 
	\begin{align}\nonumber 
	I_F(t,t', |\mathbf{p}|)
	&=
	\dfrac{\lambda}{6N} 
	\bigg\{
	\Pi_F(t, t',| \mathbf{p}|)
	- \int_0^t \diff t''\  I_\rho(t, t'', |\mathbf{p}|)
	\Pi_F(t'', t', |\mathbf{p}|)
	+ 2 \int_0^{t'} \diff  t'' \ I_F (t, t'', |\mathbf{p}|)
	\Pi_\rho ( t'', t', |\mathbf{p}|)
	\bigg\}\,, 
	\\[1ex]
	I_\rho(t,t', |\mathbf{p}|)
	&=
	\dfrac{\lambda}{3N}
	\bigg\{	 
	\Pi_\rho ( t, t', |\mathbf{p}|)
	- \int_{t'}^{t} \diff  t''  \ I_\rho (t, t'', |\mathbf{p}|)
	\Pi_\rho ( t'', t', |\mathbf{p}|)
	\bigg\}  \,, 
	\label{eq:summation_functions}%
	\end{align}%
with the one-loop terms
	\begin{align}\nonumber 
	\Pi_F(t, t', |\mathbf{p}|) 
	&=
	\int_\mathbf{q}
	\Big[
	F^\phi_{ab}(t, t', |\mathbf{p-q}|) F^\phi_{ab}(t, t', |\mathbf{q}|) 
	-\dfrac{1}{4} 	\rho^\phi_{ab}(t, t', |\mathbf{p-q}|) \rho^\phi_{ab}(t, t', |\mathbf{q}|) 
	\Big], 
	\\[1ex]
	\Pi_\rho(t,t', |\mathbf{p}|)
	&= 
	\int_\mathbf{q}
	F^\phi_{ab}(t, t', |\mathbf{p-q}|) \rho^\phi_{ab}(t, t', |\mathbf{q}|) \,.
	\end{align}
The functions $ P_F$ and $P_\rho $ describe the interactions of the quantum fluctuations with the macroscopic field. We define 
	\begin{align}\nonumber 
	H_F(t, t' , |\mathbf{p}|) 
	&=  - \sigma(t) \ F^\phi_\sigma(t, t', |\mathbf{p}|) \ \sigma(t'), \\[1ex]
	H_\rho(t, t', |\mathbf{p}|) 
	&= - \sigma(t) \ \rho^\phi_\sigma(t, t',| \mathbf{p}|)\  \sigma(t'), 
	\end{align}
in order to write down the expressions for $ P _F$ and $ P_\rho $ in momentum space as
\begin{subequations}\begin{align}
	P_F(t, t', |\mathbf{p}|)
	&= 
	- \dfrac{\lambda}{3N}
	\Bigg\{
	H_F(t, t', |\mathbf{p}|)	\nonumber\\&\qquad \qquad
	- \int_{t_0}^t \diff t'' 
	\Big[
	H_\rho(t, t'', |\mathbf{p}|) I_F(t'', t', |\mathbf{p}|)
	+ I_\rho (t, t'', |\mathbf{p}|) 
	\Big(	H_F(t'', t', |\mathbf{p}|) 
	+ J_F(t'', t'', |\mathbf{p}|)  \Big)
	\Big] \nonumber\\ & \qquad\qquad
	+ \int_{t_0}^{t'} \diff t'' 
	\Big[
	H_F(t, t'', |\mathbf{p}|) I_\rho(t'', t', |\mathbf{p}|) 
	+ I_F (t, t'', |\mathbf{p}|) 
	\Big( H_\rho(t'', t', |\mathbf{p}|)
	+ J_\rho(t'', t', |\mathbf{p}|) 	\Big)
	\Big] 	
	\Bigg\},\\
	P_\rho(t, t', |\mathbf{p}|)
	&= 
	- \dfrac{\lambda}{3N}
	\Bigg\{
	H_\rho(t, t', |\mathbf{p}|)	\nonumber\\&\qquad \qquad
	- \int_{t'}^{t} \diff t'' 
	\Big[
	H_\rho(t, t'', |\mathbf{p}|) I_\rho(t'', t', |\mathbf{p}|)
	+ I_\rho (t, t'', |\mathbf{p}|) 
	\Big(	H_\rho(t'', t', |\mathbf{p}|) 
	+ J_\rho(t'', t'', |\mathbf{p}|)  \Big)
	\Big]
	\Bigg\},
	\end{align}\end{subequations}
with the nested integrals
\begin{align}\nonumber 
	J_F(t'', t', |\mathbf{p}|) 
	&=
	\int_{t_0}^{t'} \diff s\ 
	H_F(t'', s, |\mathbf{p}|) I_\rho(s, t', |\mathbf{p}|)
	- 	
	\int_{t_0}^{t''} \diff s \ 
	H_\rho(t'', s, |\mathbf{p}|) I_F(s, t', |\mathbf{p}|), 
	\\[1ex]
	J_\rho(t'', t', |\mathbf{p}|) 
	&= 
	\int_{t''}^{t'} \diff s\ 
	H_\rho(t'', s, |\mathbf{p}|) I_\rho(s, t', |\mathbf{p}|)\,.
	\end{align}
These are the equations needed for the boson sector. Now we consider their fermionic counterpart. 
The statistical and spectral parts of the fermionic self-energy can be further decomposed into the relevant Lorentz components. With that, the statistical part of the fermion self-energy can then be expressed by the terms 
	\begin{align}\nonumber 
	C_S (x,y)&=
	- h^2 \Bigg\{
	F_S(x,y) \Big[ F_\sigma(x,y) + (N-1) F_\pi(x,y)\Big]
	- \dfrac{1}{4} \rho_S(x,y) 
	\Big[ \rho_\sigma(x,y) + (N-1) \rho_\pi(x,y)\Big]
	\Bigg\} ,\\[1ex]\nonumber 
	C^{\mu}_V (x,y)&=
	- h^2 \Bigg\{
	F^{ \mu}_V(x,y) \Big[ F_\sigma(x,y) + (N-1) F_\pi(x,y)\Big] 
	- \dfrac{1}{4} \rho^{\mu}_V(x,y) 
	\Big[ \rho_\sigma(x,y) + (N-1) \rho_\pi(x,y)\Big]
	\Bigg\},\\[1ex]
	C^{0i}_T (x,y)&=
	- h^2 \Bigg\{
	F^{0i}_T(x,y) \Big[ F_\sigma(x,y) + (N-1) F_\pi(x,y)\Big] 
	- \dfrac{1}{4} \rho^{0i}_T(x,y) 
	\Big[ \rho_\sigma(x,y) + (N-1) \rho_\pi(x,y)\Big]
	\Bigg\},
	\end{align}
while the spectral part is given by
	\begin{align}\nonumber 
	A_S (x,y)&=
	- h^2 \Bigg\{
	F_S(x,y) \Big[ \rho_\sigma(x,y) + (N-1) \rho_\pi(x,y)\Big] 
	+ \rho_S(x,y) \Big[ 
	F_\sigma(x,y) + (N-1) F_\pi(x,y)\Big]
	\Bigg\}, \\[1ex]\nonumber
	A^{\mu}_V (x,y)&=
	- h^2 \Bigg\{
	F^{\mu}_V(x,y) \Big[ \rho_\sigma(x,y) + (N-1) \rho_\pi(x,y)\Big]
	+ \rho^{\mu}_V(x,y) \Big[ 
	F_\sigma(x,y) + (N-1) F_\pi(x,y)\Big]
	\Bigg\}, \\[1ex]
	A^{0i}_T(x,y)&=
	- h^2 \Bigg\{
	F^{0i}_T(x,y) \Big[ \rho_\sigma(x,y) + (N-1) \rho_\pi(x,y)\Big] 
	+ \rho^{0i}_T(x,y) \Big[ 
	F_\sigma(x,y) + (N-1) F_\pi(x,y)\Big]
	\Bigg\}\,,
	\end{align}
where again spatial isotropy and homogeneity are implied meaning that the space-time dependence is $ (t,t', |\mathbf{x-y}|) $. With that, we have specified all the necessary terms entering the  evolution equations presented in previous section. 

\subsection{Energy-momentum tensor}
\label{sec:EMT}

Since the dynamics deduced from an effective action respect energy conservation of the system, we use energy conservation as an important indicator for the stability of our numerical simulations. We obtain the energy density from the energy-momentum tensor, which is defined as the variation of the effective action with respect to the metric $ g_{\mu\nu } (x)$,
\begin{align}
T_{\mu\nu } (x) =\left. \dfrac{2}{\sqrt{- g(x)}} \dfrac{\delta\Gamma [\phi, G, \Delta, g^{\mu\nu }]}{\delta g^{\mu\nu }} \right| _{g_{\mu\nu } = \eta _{\mu\nu }},
\end{align}
with $ -g(x)  \equiv \det g_{\mu\nu }(x)$. 
While doing so, the metric is assumed to be a general space-time-dependent metric with $ x $ being the space-time coordinate. The result is evaluated at $ g_{\mu\nu } $ equal to the Minkowski metric $ \eta _{\mu\nu } = \diag ( +1, -1, -1, -1) $. 
The energy-momentum tensor for the employed approximation scheme has been computed in \cite{tuprints2209}. Here, we provide the expressions for the energy density given by $ T_{00} $, which for a spatially homogeneous system only depends on time. Since all propagators are computed in Fourier space, we present the relevant equations in spatial momentum space, corresponding to formulas for the mode energies. These can be integrated over all momenta to obtain the energy density. 

The classical part is given by 
\begin{align}
\varepsilon^{\text{cl}}(t) =  \left[\dfrac{1}{2} \dot{\phi}^2(t) +  \dfrac{1}{2} m^2 \phi^2(t) + \dfrac{\lambda }{4!N} \phi^4(t) \right]. 
\end{align} 
The bosonic mode energy can be written as
\begin{align}\nonumber 
\varepsilon ^\phi(t, |\mathbf{p}|)
&=
\dfrac{1}{2}
\left[
\partial_t \partial_{t'} F_{aa} (t, t', |\mathbf{p}|) \Big| _{t=t'}
+ \mathbf{p}^2 F_{aa} (t, t, |\mathbf{p}|)
+ M^{2}_{\text{cl},ab} (t) F_{ba} (t, t, |\mathbf{p}|)
\right]
\\
& \quad + \dfrac{\lambda }{4!N}
F_{aa} (t, t, |\mathbf{p}|)
\int_\mathbf{q} F_{aa} (t, t, |\mathbf{q}|)
+ \dfrac{1}{2}\left[
I_F  (t, t, |\mathbf{p}|)
+ P_F  (t, t, |\mathbf{p}|)
+ \dfrac{\lambda }{3N} H_F (t, t, |\mathbf{p}|)
\right]\,,
\end{align}
while the fermionic mode energy reads
\begin{align}
\varepsilon ^\psi (t, |\mathbf{p}|)
&=
- 16 
\Big[  |\mathbf{p}| F_{V} (t, t, |\mathbf{p}|)  
+ M_\psi  (x) F_S(t, t, |\mathbf{p}|)
+ R(t, | \mathbf{p}|)
\Big]\,,
\end{align}
where 
\begin{align}
R(x^0, \mathbf{p})
&=
\quad \int_0 ^{x^0} \diff y^0
\Big[
+A_S(x^0,y^0, |\mathbf{p}|) \ F_S(y^0,x^0, |\mathbf{p}|)
- A_T(x^0,y^0, |\mathbf{p}|) \ F_T (y^0,x^0, |\mathbf{p}|)\nonumber\\
& \qquad \qquad \qquad \quad
+  A_0(x^0,y^0, |\mathbf{p}|) \ F^0_V (y^0,x^0, |\mathbf{p}|) 
- A_V(x^0,y^0, |\mathbf{p}|)  \ F_V(y^0,x^0, |\mathbf{p}|)
\Big] \nonumber\\
&\quad+ \int_0 ^{x^0} \diff y^0
\Big[
-C_S(x^0,y^0, |\mathbf{p}|) \ \rho _S(y^0,x^0, |\mathbf{p}|)
+ C_T(x^0,y^0, |\mathbf{p}|) \ \rho _T (y^0,x^0, |\mathbf{p}|)\nonumber\\
& \qquad \qquad \qquad \quad
- C_0(x^0,y^0, |\mathbf{p}|) \ \rho ^0_V (y^0,x^0, |\mathbf{p}|) 
+ C_V(x^0,y^0, |\mathbf{p}|)  \ \rho _V(y^0,x^0, |\mathbf{p}|)
\Big]\,,
\end{align}
using $ N_f = 2 $. With these expressions, the total energy density becomes 
\begin{align}
\dfrac{E(t)}{V}\equiv T_{00}(t) 
&=
\varepsilon^\mathrm{cl}(t)+ 
\int_\mathbf{p}\left[ \  \varepsilon ^\phi(t, |\mathbf{p}|)
+ \varepsilon ^\psi (t, |\mathbf{p}|) \ \right]\,. 
\end{align}
which is evaluated at each time step. 
\end{widetext}

\input{draft_qmm.bbl}
\end{document}

%% file: draft_qmm.bbl
%